%% file: paper.tex
\documentclass[useAMS,usenatbib]{mn2e}
\usepackage{mathabx}
\usepackage{longtable}
\usepackage{ulem}

\title[Satellite lines of Ne-like iron]
{Satellite lines from autoionizing states of Fe XVI and the problems with the X-ray Fe XVII lines}
\author[Del Zanna, Badnell and Storey]{G. Del Zanna$^{1}$\thanks{E-mail: gd232@cam.ac.uk},
N.~R.~Badnell$^{2}$, P.~J. Storey$^{3}$\\
$^{1}$ DAMTP, Centre for Mathematical Sciences, University of Cambridge, Wilberforce Road, Cambridge CB3 0WA, UK \\
$^{2}$ Department of Physics, University of Strathclyde,   Glasgow, G4 0NG, UK \\
$^{3}$ Department of Physics and Astronomy, University College London, London WC1E 6BT, UK \\
}
\date{Submitted to MNRAS  }

\pagerange{\pageref{firstpage}--\pageref{lastpage}} \pubyear{2023}

\DeclareMathAlphabet{\mathsc}{OT1}{cmr}{m}{sc}
\def\testbx{bx}%
\DeclareRobustCommand{\ion}[2]{%
\relax\ifmmode
\ifx\testbx\f@series
{\mathbf{#1\,\mathsc{#2}}}\else
{\mathrm{#1\,\mathsc{#2}}}\fi
\else\textup{#1\,{\mdseries\textsc{#2}}}%
\fi}

\input{def_journals}
\usepackage{graphicx,threeparttable}
\usepackage{txfonts}

\usepackage{epsfig}
\usepackage{natbib}
\usepackage{color}

\newcommand{\beq}{\begin{equation}}
\newcommand{\eeq}{\end{equation}}

\begin{document}

\label{firstpage}
\maketitle

\begin{abstract}
We present new calculations of atomic data needed to model  autoionizing states of  \ion{Fe}{xvi}.
We compare the state energies, radiative and excitation data with a sample of results
from previous literature. We find  a large scatter of results, the most significant ones
in the autoionization rates, which are very sensitive to the configuration interaction
and state mixing. We find relatively good agreement between the autoionization rates and the
collisional excitation rates calculated with the $R$-matrix suite of programs and
{\sc autostructure}. The largest model, which includes $J$-resolved states up to $n=10$, produces
ab-initio wavelengths and intensities of the satellite lines which agree well with
solar high-resolution spectra of active regions, with few minor wavelength adjustements.
We review previous literature, finding many incorrect identifications, most
notably those in the NIST database. We provide several new tentative identifications
in the 15-15.7~\AA\ range, and several new ones at shorter wavelengths, where previous
lines were unidentified. 
Compared to the previous CHIANTI model, the present one has an increased
flux  in the 15--15.7~\AA\ range at 2 MK of a factor of 1.9, resolving the discrepancies  found in 
the analysis of the Marshall Grazing Incidence X-Ray Spectrometer (MaGIXS) observation.
It appears that the satellite lines also 
resolve  the  long-standing  discrepancy  in  the intensity of the important
\ion{Fe}{xvii} 3D line at 15.26~\AA.
\end{abstract}


\section{Introduction}

The Marshall Grazing Incidence X-Ray Spectrometer (MaGIXS) flew in 2021 on
a sounding rocket and produced the first ever spectral-imaging data
of the solar corona in the X-rays, between about 6 and 30~\AA\ 
\citep{savage_etal:2023}. The instrument had a  wide slit and produced
spectroheliograms  of an X-ray bright point, which had a temperature of about 2 MK. 
The strongest emission lines in the spectra were from 
\ion{O}{vii}, \ion{O}{viii}, and the \ion{Fe}{xvii} lines between 15 and 17~\AA.
As discussed by \cite{savage_etal:2023}, the modelling of the
spectra with the CHIANTI\footnote{www.chiantidatabase.org} version 10 
atomic data \citep{chianti_v10} was satisfactory,
except the region between 15 and 15.6~\AA, where the predicted model
was lower by  nearly a factor of 2.
This is the important  spectral region where the 
strong  \ion{Fe}{xvii} resonance and intercombination lines
(3C at 15.0 and 3D at 15.26~\AA) are present. 

Possible calibration problems were excluded, which pointed to a
problem in the atomic data. 
Such a large  discrepancy was at first surprising, as $R$-matrix 
scattering calculations \citep[cf][]{hummer_etal:1993,berrington_etal:1995}
such as those of
\cite{loch_etal:06,liang_badnell:2010_ne-like}
resolved the main long-standing discrepancies between predicted and observed
intensities of the strongest \ion{Fe}{xvii} lines. Indeed, 
\cite{delzanna:2011_fe_17} showed excellent agreement, to within
10\%, between line intensities calculated with those $R$-matrix
rates and  a sample of solar high-resolution observations of
active regions and flares.
However, two of the weaker lines, the 3D at 15.26~\AA\
and the line at 15.45~\AA\ were shown  by \cite{delzanna:2011_fe_17}
to be significantly blended in quiescent  active region observations,
 where the plasma has a temperature of about 3 MK.

The discrepancies between theory and observation of the
\ion{Fe}{xvii} lines, and in particular that of the  3C/3D lines
has been the subject of well over 100 publications, many  of which are
referenced by \cite{kuhn_etal:2022}. Such interest in the literature
is because 
\ion{Fe}{xvii} provides the strongest lines in the X-rays
in laboratory and astrophysical spectra.
It is also worth noting that the strong \ion{Fe}{xvii} lines 
 can  be used to measure the electron temperature,
as confirmed  for the first time in \cite{delzanna:2011_fe_17}
[this  diagnostic was previously  known but the
earlier atomic data did not allow such a diagnostic to be used].

It has been known for a long time that satellite lines
of Ne-like iron (\ion{Fe}{xvii}), i.e. decays
to bound states from autoionizing (AI) states of Na-like \ion{Fe}{xvi} 
are present in the 14--18~\AA\ range and blend several \ion{Fe}{xvii} lines,
although a clear picture of their intensities and wavelengths has not
emerged from previous literature, as described below. 
These satellite lines are expected to be  much stronger
(relative to \ion{Fe}{xvii}) in  low-temperature 2 MK plasma.
Therefore, they are the likely candidates for the missing flux
in the MaGIXS spectra, also considering that 
the CHIANTI model for these lines was limited.

We present in this paper a selection of results from  several new atomic calculations we have
carried out and  used to calculate the intensities of these satellite lines.
The main aim of the paper is to show that indeed the missing flux in the MaGIXS spectra
is mainly due to satellite lines from AI states of Na-like \ion{Fe}{xvi}.
We also note that satellite lines from AI states of \ion{Fe}{xv} 
have also been observed in the same spectral region. 
We have carried out a preliminary calculation for the satellites from \ion{Fe}{xv}
but found them much weaker than  the \ion{Fe}{xvi} lines.  
The present and further studies are part of a long-term
programme within the  UK APAP network\footnote{www.apap-network.org}
to provide accurate atomic data for astrophysics and laboratory plasma.

The presence of the \ion{Fe}{xvi}  satellite lines needs to
be carefully taken into account when dealing with the \ion{Fe}{xvii} lines
for their diagnostic use. 
Another reason why this work is important is that the satellite lines,
once their identification and atomic data are  firmly established, could be used
for a wide range of unique diagnostic applications for solar active regions but
also in general for astrophysical plasma.  
These include measuring electron temperatures, departures from ionization equilibrium or non-Maxwellian electron distributions.
Satellite lines are usually
formed by both inner-shell (IS) excitation and dielectronic capture (DC). 
Seminal papers are \cite{gabriel_paget:1972,gabriel:1972}, while useful reviews
are those of \cite{dubau_volonte:1980,delzanna_mason:2018}.
As described in these reviews, insofar as the various diagnostics  have only been applied to
satellites of He-like ions, hence to very high temperature plasma as in solar flares. 
Therefore,  the \ion{Fe}{xvi}  satellite lines offer in principle
new diagnostic tools to study much lower temperature plasmas, typical of solar active regions.

Section~2 gives a summary of relevant previous studies.
Section~3 describes the methods and  presents
a sample of results with some comparisons with previous calculations.
Section~4 gives a sample of comparisons with solar data,
while Section~5 gives the conclusions.
A full set of atomic data in CHIANTI format is provided online via ZENODO.

\section{Earlier studies}

We now give a brief summary of the main studies on  \ion{Fe}{xvi}
autoionizing states which are relevant to the present work,
in chronological order. 
Unfortunately, none of the studies we have found in the literature
provided a complete set of data (even radiative) that we could use
to build a model for comparison with ours.
We have also tried to carry out in-depth comparisons
with the results in the literature, but very often
it has been impossible to identify states. This is not 
because of the different coupling schemes, but because the very strong
mixing within almost all the AI states means that only the  energy, parity and
$J$ could be used to try and identify states and transitions. 
The $L,S$ quantum numbers  and even the configuration are often not useful.
As {some AI states with the same parity and $J$ are very close in energy}, the ordering and mixing of states
changes considerably from calculation to calculation.
It is therefore impossible to even attempt firm comparisons with other
calculations when the full set of states is not provided.

\cite{burkhalter_etal:1979} presented laser spectra in the
15.4--16.4 and 16.8--18.0~\AA\ ranges. The spectra had a good resolution,
as they were obtained with a 3-m grazing incidence spectrometer.
The spectra contained the satellite lines from \ion{Fe}{xvi} but also
satellite lines from \ion{Fe}{xv} and many 
strong \ion{Fe}{xvii}, \ion{Fe}{xviii} transitions. 
Cowan's multi-configurational Hartree-Fock code was used to attempt the identification of several \ion{Fe}{xvi}
lines from $n=3$ states. It was clear that satellite lines are blended with many
of the strong \ion{Fe}{xvii} lines.
There were significant discrepancies between the predicted (relative) intensities
and the observed ones, as well as between the predicted and observed wavelengths.
As the authors pointed out, the identifications were very difficult, partly
because all the lines were blended, partly because the procedure  was not aided
by studies along the sequence (they also analysed similar spectra from Ti).
Despite this, several experimental energies appeared in the NIST database \citep{NIST_ASD}
from a reanalysis of the \cite{burkhalter_etal:1979} observations.   Details can be found e.g. in
\cite{shirai_etal:2000}. The NIST experimental energies were included in the
CHIANTI v.9 \citep{dere_etal:2019}  model. As we shall see below, several
of the NIST experimental energies are clearly incorrect.
Inconsistencies in the \cite{burkhalter_etal:1979} identifications were found,
although  the information was not sufficient to produce a full assessment.

\cite{jupen_etal:1988} revised a previous identification of a
single decay among the AI states, observed with the beam-foil method:
the 2p$^{5}$3s3p $^{4}$D$_{7/2}$-2p$^{5}$3s3d $^{4}$F$_{9/2}$ transition was identified with a
line they observed  at 248.36$\pm$0.05~\AA.

\cite{cornille_etal:1994} provided  a limited set of atomic data for the
$n=3$ satellite lines, calculated with the {\sc superstructure} code.
Only the total intensity factor $F_2$ (see below) was provided,
along with predicted wavelengths, and a few cross-sections for inner-shell excitation. 
Some general comparisons with SMM/FCS spectra were provided. 

\cite{phillips_etal:1997} used Cowan's code to calculate intensities
of the $n=3,4,5$  satellite lines. Although the paper was focused on
the \ion{Fe}{xvii} lines, the authors provided a table of the
strongest $n=3$ lines formed by DC, also providing a comparison with the
\cite{cornille_etal:1994} results. Unfortunately, only a few transitions
were listed, and only the total intensity factor $F_2$  was provided,
along with predicted wavelengths.  We do attempt to match the
states, and provide a comparison with our data below.
Some general comparisons with SMM/FCS spectra were also provided. 

\cite{bruch_etal:1998}  presented 
radiative data for the $n=3,4$ states calculated with 
Cowan's code and compared them
to those calculated earlier by \cite{nilsen:1989} with  YODA, a relativistic
multi-configurational Dirac-Fock code.
We refer to their tables  below when we compare our data with theirs.
Unfortunately, the authors only published (total) weighted radiative rates
and AI rates. 

\cite{brown_etal:2001} provided laboratory evidence
that at least three IS satellite lines
are present, at 15.12, 15.21, and 15.26~\AA.
The latter is blending the strong \ion{Fe}{xvii} intercombination 3D line.
They presented  low-resolution  X-ray spectra obtained with the 
Lawrence Livermore National Laboratory (LLNL) electron
beam ion trap (EBIT).

\cite{safronova_etal:2002} produced  a limited set of radiative data
for Na-like ions, calculated with their relativistic many-body codes.
Some of the energies are relatively accurate, but the data does not
include all the main $n=3$ configurations  or the strongest lines.

\cite{may_etal:2005} [M05] presented a series of laser spectra
which contained the satellite lines from AI states of
\ion{Fe}{xv} and \ion{Fe}{xvi} and  many 
strong \ion{Fe}{xvii}, \ion{Fe}{xviii} transitions. 
Most spectra had a lower resolution than those of \cite{burkhalter_etal:1979}.
\cite{may_etal:2005} 
used the Hebrew University Lawrence Livermore Atomic Code
(HULLAC) and the Flexible Atomic
Code (FAC) to present tables of intensities, wavelenghts and identifications.  
It is unclear, however,  which transitions would be relevant for astrophysical plasma
as the laser plasma produces very different spectra.

\cite{ak:2007} used GRASP to calculate radiative data 
for a limted set of  $n=3$ states. 
The energies were not as accurate as those of previous authors.

\cite{liang_etal:2008} used the $R$-matrix
suite of codes 
to calculate inner-shell electron-impact excitation (EIE) rates
of Fe$^{15+}$ for a set of $n=3$ states.
They included Auger-plus-radiation damping and showed that
earlier studies overestimated the rates.
The  \cite{liang_etal:2008} EIE and the
radiative data, calculated with {\sc autostructure} \citep[AS, see][]{badnell:2011}, were included in CHIANTI
 version 9 \citep{dere_etal:2019}. As the calculation focused on
the scattering calculations, the AS energies were not very accurate.
The CHIANTI model was complemented with AI rates  calculated with
AS, using the same set of configurations
and the same scaling parameters for the Thomas-Fermi-Amaldi
central potential, for consistency.

\cite{graf_etal:2009} [G09] presented high-resolution (about 0.05~\AA\ FWHM)
spectra obtained
in the  14.5 and 18~\AA\ range with the LLNL EBIT. The spectra
contained  the strong \ion{Fe}{xvii} and 
 inner-shell satellites from \ion{Fe}{xvi}. They used FAC
 to build an atomic model and used  the relative intensities
to provide a table of line identifications. No details on the atomic calculations or data
 were given. 
 They also produced calculated spectra from the \cite{cornille_etal:1994}
 and \cite{phillips_etal:1997} data in a way that was not described, and  concluded
 that the wavelenghts and relative intensities based on  those  previous studies
 were completely wrong.
 However, even the comparisons with the FAC model spectra was not entirely
 satisfactory and complicated by blending with many transitions.

\cite{diaz_etal:2013} [D13]  produced a set of accurate energies for the $n=3$ states,
calculated with the relativistic Multireference Moller–Plesset (MR-MP) perturbation theory.
For a selection of transitions, they provided wavelengths and a comparison to those
calculated by \cite{may_etal:2005} with HULLAC. They also report a table of identifications
presumably based on wavelength coincidences with the observations reported by
\cite{graf_etal:2009}. 
As \cite{diaz_etal:2013} provided the full set of energies, it was possible to
identify the correspondence with our calculations in most cases.

\cite{beiersdorfer_etal:2012} used the \cite{diaz_etal:2013}
energies and a  set of unpublished FAC calculations to
revise several previous identifications of inner-shell satellites suggested by 
\cite{graf_etal:2009}. The authors also attempted to identify IS lines
in a Chandra spectrum of Capella, although in that spectrum the
lower-temperature lines are very weak, and lines from ions hotter
than \ion{Fe}{xvii}  are also present. 

In a follow-up paper, 
\cite{beiersdorfer_etal:2014} used the \cite{diaz_etal:2013}
energies and a  set of unpublished FAC calculations to indicate the
predicted wavelengths of the strongest satellites formed by DC,
against the Chandra spectrum of Capella. 
In a similar study, more extended FAC calculations (up to $n=30$) were used
by \cite{beiersdorfer_etal:2018} to
predict the intensities of $n\ge 4$ lines formed by DC.
No details were provided, although a
plot in \cite{beiersdorfer_etal:2011} on what is presumably the
same calculation  shows the various
contributions from AI states, from which it appears that
nearly  all the flux blending the 3C line comes from AI states
between $n=4$ and $n=9$.


\section{Methods}

Considering only dielectronic capture, 
the population $N_s$ of the autoionising state $s$ of the Na-like iron is
 determined by the 
balance between the dielectronic capture (with rate $C^{\rm dc}$),
autoionisation and radiative decay  to all energetically lower levels:
\beq
N_{\rm Ne-like}  N_{\rm e} C^{\rm dc} = N_s  \left( \sum_k A^{\rm a}_{sk} + \sum_{f<s} A_{sf} \right)
\eeq
where  $A_{sf}$ is the transition probability to a decay to a bound state (decay rate for short),
$ A^{\rm a}_{sk}$ is the autoionisation rate, $N_{\rm e}$ the electron number density,
and $N_{\mathrm{Ne-like}}$ the ground-state population of the Ne-like iron. 
The intensity of the satellite line decay from the state $s$ to the bound
state $f$ is therefore proportional to 
\beq
I^{\rm dc}_{sf} =  N_{\rm Ne-like}  N_{\rm e} C^{\rm dc}  {A_{sf} \over \sum_k A^{\rm a}_{sk} + \sum_{f<s} A_{sf} } \,.
\eeq
By applying the Saha equation for thermodynamic equilibrium
we obtain a  relation between  dielectronic capture and autoionisation rates
and find that $C^{\rm dc} \propto g_s \, \sum_k A^{\rm a}_{sk}$, where $g_s$ is the 
statistical weight of the autoionizing state.
The intensity of the spectral line is therefore proportional to the factor $F_2$:
\beq
F_2 = {g_s \, A_{sf} \, \sum_k A^{\rm a}_{sk} \over \sum_k A^{\rm a}_{sk} + \sum_{f<s} A_{sf} } \, = g_s \, A_{sf} \, Y \quad [s^{-1}] \,
\eeq
which for strong lines is of the order of 10$^{13}$ s$^{-1}$ or higher.
We provide below Tables of these factors $F_2$, for comparison to earlier
literature, when available. We also list the ratio $Y$, which is an indication of how close the
state is to LTE: when the AI rate $ \sum_k A^{\rm a}_{sk}$ is much larger than the
decay rate, $Y \simeq 1$ and the uncertainty in the AI rate does not have a
signifcant effect on the line intensity.

\begin{table}
\caption{ The target electron configuration basis 
 and orbital scaling parameters $\lambda_{nl}$ 
for the structure  run of the  $n=3$ model. 
}
\begin{flushleft}
\begin{tabular}{l|llll}
\hline\hline\noalign{\smallskip}
Configurations     &               &  &    &  \\ 
\noalign{\smallskip}\hline\noalign{\smallskip}
 1s$^2$ 2s$^2$ 2p$^6$ 3$l$  ($l$=s,p,d)          &   1s & 1.39933 &     &  \\
 1s$^2$ 2s$^2$ 2p$^5$ 3$l$ 3$l'$  ($l,l'$=s,p,d) &   2s & 1.15549 & 3s  & 1.13314 \\
1s$^2$ 2s  2p$^6$ 3$l$ 3$l'$  ($l,l'$=s,p,d)     &  2p & 1.09545 & 3p  &  1.09941  \\ 
                                                 & 3d &  1.13123 &    & \\ 
 \noalign{\smallskip}\hline
\end{tabular}
\normalsize
\end{flushleft}
\label{tab:basis3}
\end{table}

Clearly, the actual line intensity does not scale linearly with the factors $F_2$
if the inner-shell excitation is significant.
As the Na-like iron does not have metastable states for
coronal densities, inner-shell excitation can only be a significant  populating process 
for strong decays to the ground state.
Generally, the intensity of the satellite line depends both on
dielectronic capture and inner-shell.
Our approach is to obtain the  intensities of the satellite lines by solving
the collisional-radiative matrix which includes both the Na- and Ne-like ions,
using the IDL codes developed by one of us (GDZ) and made available
to the community via CHIANTI version 9, as described in the Appendix of the
paper.
We adopt the total dielectronic recombination (DR) rate coefficients between the
ground states of the Ne- and Na-like iron from the UK APAP network.
We also include  in our model level-resolved radiative recombination, with the data
also from the  UK APAP network.

\begin{table*}
\caption{ The target electron configuration basis 
 and orbital scaling parameters $\lambda_{nl}$ 
for the structure and DW runs of the $n=6$ model. 
}
\begin{flushleft}
\begin{tabular}{l|llllllllllllllll}
\hline\hline\noalign{\smallskip}
Configurations     &               &  &     &  & & & & & &   \\ 
\noalign{\smallskip}\hline\noalign{\smallskip}
 1s$^2$ 2s$^2$ 2p$^6$ 3$l$  ($l$=s,p,d)       &   1s &  1.39863 &     &  & & & & & &  \\
 1s$^2$ 2s$^2$ 2p$^6$ 4$l$  ($l$=s,p,d,f)     &   2s &  1.15549 & 3s  & 1.13323 & 4s &   1.13105 & 5s & 1.12566 & 6s & 1.12792 \\
 1s$^2$ 2s$^2$ 2p$^6$ 5$l$  ($l$=s,p,d,f,g)    &  2p &  1.09540 & 3p  & 1.09919 & 4p &   1.09824 & 5p & 1.09322 & 6p & 1.09651 \\ 
 1s$^2$ 2s$^2$ 2p$^6$ 6$l$  ($l$=s,p,d,f,g,h)               & 3d & 1.13139 & 4d & 1.12449 & 5d & 1.11611  & 6d & 1.11910 &  & \\
 1s$^2$ 2s$^2$ 2p$^5$ 3$l$ $nl'$  ($l$=s,p,d, $n$=3-6,$l'$=s,p,d,f,g,h)  & 4f & 1.22747 & 5f & 1.19739 & 6f & 1.20392  & & & & \\
1s$^2$ 2s  2p$^6$ 3$l$ $nl'$  ($l$=s,p,d, $n$=3-6,$l'$=s,p,d,f,g,h)      & 5g & 1.33528 & 6g & 1.35001 & 6h & 1.0 & & & &  \\
 \noalign{\smallskip}\hline
\end{tabular}
\normalsize
\end{flushleft}
\label{tab:basis6}
\end{table*}

\begin{table*}
\caption{ The target electron configuration basis 
 and orbital scaling parameters $\lambda_{nl}$ 
 for the structure and DW runs of the $n=10$ model.
}
\begin{flushleft}
\begin{tabular}{l|llllllllllllllll}
\hline\hline\noalign{\smallskip}
Configurations     &               &  &     &  & & & & & &   \\ 
\noalign{\smallskip}\hline\noalign{\smallskip}
 1s$^2$ 2s$^2$ 2p$^6$ 3$l$  ($l$=s,p,d)       &   1s &  1.40272 &     &  & & & & & &  \\
 1s$^2$ 2s$^2$ 2p$^6$ 4$l$  ($l$=s,p,d,f)     &   2s &  1.14452 & 3s  & 1.15069 & 4s &   1.12 & 5s & 1.12052 & 6s & 1.12065 \\
 1s$^2$ 2s$^2$ 2p$^6$ 5$l$  ($l$=s,p,d,f,g)    &  2p &  1.08489 & 3p  & 1.08734 & 4p &   1.08916 & 5p & 1.08936 & 6p & 1.08962 \\ 
 1s$^2$ 2s$^2$ 2p$^6$ $nl$  ($n=6-10,l$=s,p,d,f,g,h)  & 3d &  1.10032 & 4d & 1.11 & 5d & 1.11005  & 6d & 1.11135 &  & \\
 1s$^2$ 2s$^2$ 2p$^5$ 3$l$ $nl'$ ($l$=s,p,d, $n$=3-10,$l'$=s,p,d,f,g,h)  & 4f & 1.18687 & 5f & 1.18174 & 6f & 1.18411  & & & & \\
1s$^2$ 2s  2p$^6$ 3$l$ $nl'$  ($l$=s,p,d, $n$=3-10,$l'$=s,p,d,f,g,h)      & 5g & 1.31015 & 6g & 1.32363 & 6h & 1.10890 & & & &  \\
 \noalign{\smallskip}\hline
\end{tabular}
\normalsize
\end{flushleft}
\label{tab:basis10}
\end{table*}

We do not attept to model the high-density laser spectra, for a number of reasons.
First, many excited states become populated and CE rates need to be included.
Second, level-resolved recombination needs to be included in the model.
Third, the treatment of the DR process does not include transitions among the
AI states (collisional redistribution before they can relax via
radiation or autoionisation), which become non-negligible for high-density plasma.
Fourth, continuum lowering also needs to be modelled.
Fifth, plasma conditions are such that non-Maxwellian electron distributions (NMED)
and time-dependent effects naturally arise in the plasma.
Sixth, modelling the relative abundance of the Ne- and Na-like ions is non-trivial.
Seventh, coupling with the background radiation field in the level population modelling
should be taken into account.

The radiative data,  $A_{sf}$ and   $A^{\rm a}_{sk}$, has been calculated with
{\sc autostructure}.
We have run many calculations by increasing the number of configurations,
as described below.

{\sc autostructure} has a very large set of parameters and ways to run a calculation,
in other words is extremely flexible. 
We experimented with different potentials. We tried a new development, which includes
the same  potential and optimization parameters as used by FAC, but it did not improve the results. 
We also tried semi-relaxed orbitals, where groups of configurations each
have their own potential scaling parameters; and also the fully-relaxed case, where each configuration (initially) uses its own Slater-Type-Orbital potential built from its occupation numbers and, optionally, this can be iterated to self-consistency. The fully-relaxed case produced excellent energies 
for the lowest set of $n=3$ configurations, but diverged by nearly 10,000 cm$^{-1}$
for the highest ones. 
At the end we used  a unique set of orbitals with Thomas--Fermi scaling
parameters optimized by minimizing an energy functional
which included first the terms from the $n=2,3$ configurations, and then
iteratively those arising from higher shells.

For the calculation of the $A^{\rm a}_{sk}$ rates, we included the 
6 lowest excited states in the Na-like ion.
Finally, we found that the use of the kappa-averaged semi-relativistic potential
improves  the results.
We also added two-body non-fine-structure
      interactions (contact spin-spin, two-body Darwin and
      orbit-orbit),  Breit and QED corrections.

\subsection{CE rates}

We have complemented each set of radiative data with {\sc autostructure}
Breit-Pauli distorted wave (DW) calculations with the same target.
We included excitations from the ground state as well as the first
four excited states, although for most astrophysical applications
the population of Na-like iron is all in the ground state.
The rates from excited states have been included in the models
for a simple assessment of how different the relative intensities
of the lines are for high-density plasmas.

Collision strengths are calculated at
the same set of final scattered energies for all transitions.
`Top-up' for the contribution of high partial waves is done using the same Breit-Pauli
methods and subroutines implemented in the R-matrix outer-region code STGF.
The collision strengths were extended to high
 energies by interpolation using  the appropriate high-energy limits,
 while the  temperature-dependent effective 
collisions strength $\Upsilon(i-j)$ (CE rate coefficients) were calculated by assuming a 
Maxwellian electron distribution and linear integration 
with the final energy of the colliding electron.

\subsection{Other rates}

When building the collisional-radiative models, 
 we used the $R$-matrix CE rate coefficients and radiative data of 
\cite{liang_etal:2008} for the bound states included in their model.
For the Ne-like ion, we adopted the CHIANTI version 10 \citep{chianti_v10} model.


\section{A sample of results}

We have run many AS  calculations. We started 
 with the $n=3$ set shown in Table~\ref{tab:basis3}.
 This set is more complete  compared to that of \cite{liang_etal:2008}
 (shown in Appendix), as it includes
the 2s$^{2}$ 2p$^{5}$ 3d$^{2}$ configuration, not present in the
earlier calculation, but that produces strong satellite lines via dielectronic capture.
We have also added configurations with double excitations opening the 2s$^{2}$ shell.
We kept the 1s$^{2}$ shell closed.
The corresponding bound states were included.

We then increased the size by adding the $n=4$, $n=5$, and $n=6$ orbitals and
corresponding set of configurations. The $n=6$ model has 120 configurations and
3450 fine-structure levels.
Table~\ref{tab:basis6} lists the set of configurations and scaling parameters
adopted for this  $n=6$  model. 

For each run we compared the energies with the experimental ones from NIST
and the \cite{diaz_etal:2013} thoretical energies. 
We also experimented with opening the 1s shell and triple excitations, but the
energies did not improve much.


The $n=6$  model provided thoretical energies  very close to
\cite{diaz_etal:2013} and was a baseline to try and identify
spectral lines using various observations, and to compare with
previous identifications when possible. It provided predictions
for all the observable lines.  
With few exceptions, relatively good agreement with observations
was found.

However, considering the \cite{beiersdorfer_etal:2011} results,
a further calculation was carried out, including all the
main configurations up to $n=10$, to improve the
predictions for the series of satellite lines blending the 3C resonance lines.
The size of the model is
large (288 configurations for 8886 j-resolved states)
but still feasible with the CHIANTI programs.
Table~\ref{tab:basis10} lists the set of configurations and scaling parameters $\lambda_{nl}$
adopted for this  $n=10$  model. 
The $\lambda_{nl}$ for  $n=7-10$ have been kept equal to the $n=6$ ones, as it is clear that
they vary little with $n$. 
It turns out that
the ab-initio wavelengths are better than the $n=6$  model,
although for most transitions up to $n=5$ the resulting intensities are
close to those of the  $n=6$  model.

Finally, to have an estimate of the contributions
from even  higher configurations, we have carried out a
configuration-averaged AS calculation including the same set of
configurations up to $n=30$.

\subsection{Energies}

\begin{table}
\begin{center}
  \caption{List of the main states.
    \label{tab:table_e_n6} }
\footnotesize
\setlength\tabcolsep{3.5pt} 
\begin{tabular}{@{}rllllllllllllllllllllll@{}}
  \hline\noalign{\smallskip}
 $i$ &  Conf.  & P & T & $E_{\rm exp}$ &  $E_{\rm AS}$  & $E_{\rm Diaz+}$ & $E_{\rm Liang+}$  \\ 
\noalign{\smallskip}
  \hline
    1 &         2s$^{2}$ 2p$^{6}$ 3s &   e &   $^2$S$_{1/2}$ &         0 &         0 &         0 &         0 & \\ 
    2 &         2s$^{2}$ 2p$^{6}$ 3p &   o &   $^2$P$_{1/2}$ &    277194 &    277711 &    277222 &    276436 & \\ 
    3 &         2s$^{2}$ 2p$^{6}$ 3p &   o &   $^2$P$_{3/2}$ &    298143 &    300089 &    298167 &    296534 & \\ 
    4 &         2s$^{2}$ 2p$^{6}$ 3d &   e &   $^2$D$_{3/2}$ &    675501 &    676579 &    675463 &    676373 & \\ 
    5 &         2s$^{2}$ 2p$^{6}$ 3d &   e &   $^2$D$_{5/2}$ &    678405 &    681330 &    678372 &    679712 & \\ 
\noalign{\smallskip}
   33 &   2s$^{2}$ 2p$^{5}$ 3s$^{2}$ &   o &   $^2$P$_{3/2}$ &   5773000? &   5744641 &   5756556 &   5802584 & \\ 
   34 &   2s$^{2}$ 2p$^{5}$ 3s$^{2}$ &   o &   $^2$P$_{1/2}$ &   5873000? &   5848114 &   5857665 &   5899697 & \\ 
   35 &      2s$^{2}$ 2p$^{5}$ 3s 3p &   e &   $^4$S$_{3/2}$ &  -   &   5939043 &   5953391 &   5991935 & \\ 
   36 &      2s$^{2}$ 2p$^{5}$ 3s 3p &   e &   $^4$D$_{5/2}$ &   5982000 &   5967095 &   5980479 &   6020272 & \\ 
   37 &      2s$^{2}$ 2p$^{5}$ 3s 3p &   e &   $^4$D$_{7/2}$ &  -   &   5973428 &   5986775 &   6026148 & \\ 
   38 &      2s$^{2}$ 2p$^{5}$ 3s 3p &   e &   $^2$P$_{3/2}$ &  -   &   5974184 &   5987047 &   6027021 & \\ 
   39 &      2s$^{2}$ 2p$^{5}$ 3s 3p &   e &   $^2$P$_{1/2}$ &   6001000 &   5986456 &   5999543 &   6041011 & \\ 
   40 &      2s$^{2}$ 2p$^{5}$ 3s 3p &   e &   $^4$P$_{5/2}$ &   6013000 &   5998400 &   6011855 &   6053544 & \\ 
   41 &      2s$^{2}$ 2p$^{5}$ 3s 3p &   e &   $^2$D$_{3/2}$ &   6013000 &   5999767 &   6012375 &   6053898 & \\ 
   42 &      2s$^{2}$ 2p$^{5}$ 3s 3p &   e &   $^2$S$_{1/2}$ &   6042000? &   6016544 &   6027754 &   6076536 & \\ 
   43 &      2s$^{2}$ 2p$^{5}$ 3s 3p &   e &   $^4$D$_{1/2}$ &   6075000 &   6066510 &   6077192 &   6113566 & \\ 
   44 &      2s$^{2}$ 2p$^{5}$ 3s 3p &   e &   $^4$P$_{1/2}$ &   6089000? &   6072233 &   6082835 &   6128206 & \\ 
   45 &      2s$^{2}$ 2p$^{5}$ 3s 3p &   e &   $^4$D$_{3/2}$ &   6089000 &   6077288 &   6087509 &   6124285 & \\ 
   46 &      2s$^{2}$ 2p$^{5}$ 3s 3p &   e &   $^2$D$_{5/2}$ &       -   &   6087412 &   6096282 &   6141431 & \\ 
   47 &      2s$^{2}$ 2p$^{5}$ 3s 3p &   e &   $^4$P$_{3/2}$ &   6096000 &   6089389 &   6100268 &   6138528 & \\ 
   48 &      2s$^{2}$ 2p$^{5}$ 3s 3p &   e &   $^2$D$_{5/2}$ &   6110000 &   6098693 &   6108077 &   6147237 & \\ 
   49 &      2s$^{2}$ 2p$^{5}$ 3s 3p &   e &   $^2$P$_{3/2}$ &   6129000? &   6105380 &   6113831 &   6157761 & \\ 
   50 &      2s$^{2}$ 2p$^{5}$ 3s 3p &   e &   $^2$P$_{1/2}$ &  -   &   6178837 &   6182346 &   6229457 & \\ 
   51 &      2s$^{2}$ 2p$^{5}$ 3s 3p &   e &   $^2$D$_{3/2}$ &   6217000? &   6195854 &   6201702 &   6244142 & \\ 
   52 &      2s$^{2}$ 2p$^{5}$ 3s 3p &   e &   $^2$S$_{1/2}$ &   6267000? &   6252510 &   6245187 &   6313279 & \\

\noalign{\smallskip}
   67 &      2s$^{2}$ 2p$^{5}$ 3s 3d &   o &   $^4$P$_{5/2}$ &   6393000 &   6379612 &   6390567 &   6440048 & \\ 
\noalign{\smallskip}
   73 &      2s$^{2}$ 2p$^{5}$ 3s 3d &   o &   $^4$F$_{5/2}$ &   6406000 &   6394429 &   6404701 &   6453145 & \\ 
   74 &   2s$^{2}$ 2p$^{5}$ 3p$^{2}$ &   o &   $^2$P$_{3/2}$ &      -   &   6397961 &   6406003 &   6469670 & \\ 
   75 &      2s$^{2}$ 2p$^{5}$ 3s 3d &   o &   $^2$D$_{3/2}$ &   6419000 &   6405385 &   6415660 &   6464402 & \\ 
   76 &   2s$^{2}$ 2p$^{5}$ 3p$^{2}$ &   o &   $^2$D$_{3/2}$ &  -   &   6413008 &   6422064 &   6455698 & \\ 
   77 &      2s$^{2}$ 2p$^{5}$ 3s 3d &   o &   $^4$D$_{7/2}$ &   6422000 &   6413123 &   6421329 &   6471649 & \\ 
   78 &      2s$^{2}$ 2p$^{5}$ 3s 3d &   o &   $^2$P$_{1/2}$ &  6423000  &   6415027 &   6423498 &   6464730 & \\ 
   79 &      2s$^{2}$ 2p$^{5}$ 3s 3d &   o &   $^2$F$_{5/2}$ &  6425000  &   6415181 &   6423578 &   6476262 & \\
   80 &   2s$^{2}$ 2p$^{5}$ 3p$^{2}$ &   o &   $^2$D$_{5/2}$ &  -  &   6417371 &   6425339 &   6474699 & \\
   81 &      2s$^{2}$ 2p$^{5}$ 3s 3d &   o &   $^2$P$_{3/2}$ &  \sout{6436000}  &   6436676 &   6443091 &   6498398 & \\
      & & & & {\bf 6444100?} & & & \\
   82 &      2s$^{2}$ 2p$^{5}$ 3s 3d &   o &   $^4$D$_{1/2}$ &              -   &   6447867 &   6455202 &   6506414 & \\ 
   83 &      2s$^{2}$ 2p$^{5}$ 3s 3d &   o &   $^4$D$_{3/2}$ &  \sout{6473000}   &   6476481 &   6483365 &   6536053 & \\ 
   84 &      2s$^{2}$ 2p$^{5}$ 3s 3d &   o &   $^2$F$_{7/2}$ &  \sout{6445000}   &   6480673 &   6485011 &   6546990 & \\ 
   85 &      2s$^{2}$ 2p$^{5}$ 3s 3d &   o &   $^4$F$_{3/2}$ &   6502000 &   6493786 &   6502061 &   6547383 & \\ 
   86 &      2s$^{2}$ 2p$^{5}$ 3s 3d &   o &   $^2$D$_{5/2}$ &  \sout{6464000}   &   6495941 &   6501608 &   6549706 & \\ 
   87 &      2s$^{2}$ 2p$^{5}$ 3s 3d &   o &   $^4$D$_{5/2}$ &   6502000 &   6496993 &   6504077 &   6555370 & \\
\noalign{\smallskip}
\hline
\end{tabular}
\begin{tablenotes}
    \item[] { $E_{\rm exp}$ gives the NIST experimental energies, except the
   those in bold which are our tentative values. $E_{\rm AS}$ are our ab-initio
   AS energies with the $n=6$ model. $E_{\rm Diaz+}$ are the energies from \cite{diaz_etal:2013}
while  $E_{\rm Liang+}$ are the AS ones from \cite{liang_etal:2008}. }
 \end{tablenotes}
\end{center}
\normalsize
\end{table}

\addtocounter{table}{-1}
\begin{table}
\begin{center}
   \caption{Contd  }
\footnotesize
\setlength\tabcolsep{3.5pt} 
\begin{tabular}{@{}rllllllllllllllllllllll@{}}
  \hline\noalign{\smallskip}
 $i$ &  Conf.  & P & T & $E_{\rm NIST}$ &  $E_{\rm AS}$  & $E_{\rm Diaz+}$ & $E_{\rm Liang+}$  \\ 
\noalign{\smallskip}
  \hline
   88 &   2s$^{2}$ 2p$^{5}$ 3p$^{2}$ &   o &   $^2$P$_{1/2}$ &  -   &   6504938 &   6508883 &   6566725 & \\ 
   89 &      2s$^{2}$ 2p$^{5}$ 3s 3d &   o &   $^2$D$_{5/2}$ &   6516000 &   6508973 &   6514575 &   6569938 & \\ 
   90 &      2s$^{2}$ 2p$^{5}$ 3s 3d &   o &   $^2$F$_{7/2}$ &   6517000 &   6509407 &   6514871 &   6561652 & \\ 
   91 &   2s$^{2}$ 2p$^{5}$ 3p$^{2}$ &   o &   $^2$P$_{1/2}$ &  -   &   6511585 &   6514341 &   6579688 & \\ 
   92 &   2s$^{2}$ 2p$^{5}$ 3p$^{2}$ &   o &   $^2$P$_{3/2}$ &  -   &   6528784 &   6531608 &   6592253 & \\ 
   93 &      2s$^{2}$ 2p$^{5}$ 3s 3d &   o &   $^2$D$_{3/2}$ &   \sout{6530000} &   6549199 &   6550184 &   6611638 & \\
        &                             &     &                &  {\bf 6553500?}   &        &        &      \\ 

   94 &      2s$^{2}$ 2p$^{5}$ 3s 3d &   o &   $^2$P$_{1/2}$ &   6574000 &   6575409 &   6573657 &   6644694 & \\ 
   95 &      2s$^{2}$ 2p$^{5}$ 3p 3d &   e &   $^4$D$_{1/2}$ &  -   &   6586884 &   6601400 &   6646599 & \\ 
   96 &      2s$^{2}$ 2p$^{5}$ 3s 3d &   o &   $^2$F$_{5/2}$ &   \sout{6556000}   &   6591507 &   6593543 &   6651977 & \\ 
   97 &      2s$^{2}$ 2p$^{5}$ 3p 3d &   e &   $^4$D$_{3/2}$ &  -   &   6595119 &   6608991 &   6654887 & \\ 
   98 &      2s$^{2}$ 2p$^{5}$ 3p 3d &   e &   $^4$D$_{5/2}$ &  -   &   6607993 &   6620899 &   6667902 & \\ 
   99 &      2s$^{2}$ 2p$^{5}$ 3s 3d &   o &   $^2$P$_{3/2}$ &   \sout{6595000} &   6617260 &   6616740 &   6686516 & \\
      &                             &     &                &   {\bf 6620000?} &        &        &      \\
\noalign{\smallskip}
  151 &      2s$^{2}$ 2p$^{5}$ 3p 3d &   e &   $^2$D$_{3/2}$ &   {\bf 6831000?} &   6833056 &   6831282 &   6894915 & \\ 
  152 &      2s$^{2}$ 2p$^{5}$ 3p 3d &   e &   $^2$D$_{5/2}$ &   {\bf 6837100?} &   6837436 &   6838045 &   6893483 & \\

\noalign{\smallskip}
  191 &   2s$^{2}$ 2p$^{5}$ 3d$^{2}$ &   o &   $^2$G$_{7/2}$ &   {\bf  7135000?}  &   7130233 &   7134361 &  -   & \\ 
  192 &   2s$^{2}$ 2p$^{5}$ 3d$^{2}$ &   o &   $^4$F$_{5/2}$ &  -   &   7136863 &   7142053 &  -   & \\ 
193 &            2s 2p$^{6}$ 3s 3p &   o &   $^2$P$_{3/2}$ &  -   &   7138849 &   7128949 &  -   & \\

\noalign{\smallskip}
  200 &   2s$^{2}$ 2p$^{5}$ 3d$^{2}$ &   o &   $^2$F$_{5/2}$ &  {\bf  7180800?} &   7183602 &   7186088 &  -   & \\ 
  201 &   2s$^{2}$ 2p$^{5}$ 3d$^{2}$ &   o &   $^2$F$_{5/2}$ &  {\bf  7191000?} &   7188617 &   7193832 &  -   & \\ 

\noalign{\smallskip}
  210 &   2s$^{2}$ 2p$^{5}$ 3d$^{2}$ &   o &   $^2$F$_{7/2}$ &  {\bf 7236000?} &   7241831 &   7242818 &  -   & \\ 
  211 &   2s$^{2}$ 2p$^{5}$ 3d$^{2}$ &   o &   $^2$D$_{5/2}$ &  {\bf 7240000?} &   7251131 &   7246119 &  -   & \\ 
  212 &   2s$^{2}$ 2p$^{5}$ 3d$^{2}$ &   o &   $^2$D$_{3/2}$ &  -   &   7257952 &   7253388 &  -   & \\ 
  213 &   2s$^{2}$ 2p$^{5}$ 3d$^{2}$ &   o &   $^2$P$_{3/2}$ &  {\bf 7266000?} &   7269242 &   7263947 &  -   & \\ 

\noalign{\smallskip}
\hline
\end{tabular}
\end{center}
\normalsize
\end{table}

Table~\ref{tab:table_e_n6} lists the energies of a few  $n=3$ bound states
and those of a selection of AI $n=3$ states.  The configuration and
LS labelling is from AS but is often not very meaningful. 
The first column lists the  experimental energies which
are from NIST, except a few new tentative identifications,
while the second column  gives
the AS values obtained with the  $n=6$ model. The following two
columns list the \cite{diaz_etal:2013} and \cite{liang_etal:2008} values.
Table~\ref{tab:table_e_n10} in the Appendix lists the energies as obtained 
with the  $n=10$ model, which illustrates how little the values change with the size
of the calculation.
A full comparison with  the \cite{diaz_etal:2013} energies is provided in the Appendix.


Our AS energies are generally very close (within a few thousands
of cm$^{-1}$) to the \cite{diaz_etal:2013} ones, especially
for the states producing the strongest solar spectral lines. Note that
the uncertainty in the experimental values has a similar magnitude.
The comparisons with the experimental energies we have carried out indicate
that the \cite{diaz_etal:2013}  are the most accurate
energies across the literature. This is one of the reasons why we have included them
in the Table. The other is that the authors provided the full set of
states so we could match them against ours with some confidence.

Our energies predict the 2p$^{5}$3s3p $^{4}$D$_{7/2}$
(level No.81) -2p$^{5}$3s3d $^{4}$F$_{9/2}$ (level No.112) transition, identified with a
line at 248.36$\pm$0.05~\AA\ by 
\cite{jupen_etal:1988}, to be at 245.9~\AA, while the
 \cite{diaz_etal:2013} energies predict 248.5~\AA.

Table~\ref{tab:table_e_n6} also clearly shows that the 
\cite{liang_etal:2008} energies differ by a significant amount,
about 40,000 cm$^{-1}$. This is the reason why the mixing of the states
and ultimately the rates obtained with  the \cite{liang_etal:2008} model
are sometimes quite different than those we have calculated, as shown below.

Table~\ref{tab:table_e_n6} also shows that in several cases the
NIST energies must be incorrect, not only because of the large
departures from the \cite{diaz_etal:2013}  (or our) values, but also because the intensities 
of their decays do not match solar observations, as briefly outlined below.
We have highlighted the main
ones, but in several other cases, where
a question mark is added, the NIST values might also be wrong.

Unfortunately, we have a circular problem: the published  structure calculations
provide different wavelengths and intensities, hence different
identifications. 
As pointed out by \cite{burkhalter_etal:1979},
studies along the sequence do not help.
For a few states producing the strongest lines,
we have provided tentative new energies and used them for the comparisons
to observations. Details are provided below.
We have not attempted to apply semi-empirical corrections to the 
calculations as implemented within AS, and which would improve the results.
They could be applied in the future,
once the main transitions will be firmly established
with new laboratory and solar spectra.

\subsection{Radiative data }

Table~\ref{tab:table2} shows as an example 
the CHIANTI radiative data, which were obtained from the \cite{liang_etal:2008} $n=3$
target and those we have calculated with the more extended $n=3$ sets of configurations
and different scaling parameters of Table~1. 
Only transitions from the lowest AI states and with an intensity factor $F_2$ larger than 5x10$^{11}$ 
are shown. Note that typical values of $F_2$ of observed lines in astrophysical spectra are
10$^{13}$ or higher, although the large number of weak transitions within short
wavelength intervals means that weaker transitions can also be significant. 
We have removed from the model transitions with branching ratios less than
10$^{-5}$ but the total number of satellite lines, all within 11--18~\AA, is over 700 000.

We can see that differences of a factor of two in the  $F_2$ values are common,
although some transitions, indicated in the last column, are actually 
mainly formed by IS, and not DC, hence the $F_2$ value is not related to the actual
intensity of the line. Sometimes the differences are related to the
decay rate, sometimes with the AI rate. 

There are also cases as the first transition in the Table where the decay rate
is similar but the AI rate is different by nearly a factor of three.
By running many calculations, we found that even small changes in the CI
expansion or the scaling parameters can have a large effect on the AI
rates, easily by an order of magnitude. On the other hand, the
radiative rates are generally less affected.
Almost all the AI states are completely mixed,
and any small change in the relative energies can have a large effect on the mixing
and on the AI rates.
The fact that the radiative data are often less affected is due to the
different sensitivity to the short and long-range parts of the wavefunctions.

We are not aware that this important issue has been highlighted in the literature.
On the other hand, it is also worth pointing out that a large uncertainty in the
AI rate does not  affect the line intensity when the ratio $Y$ is close
to unity, as the AI state is in LTE (AI rate dominant over the decay
rate). We have highlighted in the last column the cases where $Y$ is
much lower than 1 and the calculated AI rates vary significantly.

To validate the AS AI rates,  we have run a calculation with the 
$n=3$ set,  switched off the corrections for the two-body non-fine-structure
      interactions, 
and run the Breit-Pauli $R$-matrix (BPRM) suite of codes with a relatively 
simple Ne-like target (4 configurations). We  used the
Quigley and Berrington method \citep{qb:1996,qb:1998} to locate resonances and get their widths.
This process is  time consuming, so only a sample of values are shown. 
 We are not aware of any such comparison presented in the literature.
Table~\ref{tab:table2}  shows the ratio $R$ between the AS
AI rates and those  calculated  from the widths of the resonances.

The comparison with the $R$-matrix AI rates is reassuring, with typical
differences for the stronger transitions of 10--30\%.
However, in a few cases large differences are present. 
We have verified that they occur when two states that are mixing are very close in energy.
Generally, the $R$-matrix energies are quite different from the
AS values, even using the same target, so the mixing of states is often
quite different.

\begin{table*}
\begin{center}
   \caption{List of the main transitions from the lowest states formed by dielectronic capture ($n$=3 models).
   F2 values, as well as  $A_{ji}$ and AI rates are in 10$^{13}$ s$^{-1}$. 
\label{tab:table2} }
\footnotesize
\setlength\tabcolsep{4.pt} 
\begin{tabular}{@{}rllllllllllllllllllllll@{}}
  \hline\noalign{\smallskip}
 $j$ & $i$&      C$_j$         & T$_j$        & C$_i$  & T$_i$     & $\lambda$ & $F_2$  & $A_{ji}$ & $F_2$ & $Y$   & $A_{ji}$& AI & AI  & R & \\
     &     &                   &             &        &           &    \AA    & v9  &  V9     &$n=3$&$n=3$ & $n=3$  & v9 & $n=3$& & \\
\noalign{\smallskip}
  \hline
  33 &  1 &    2p$^{5}$ 3s$^{2}$ & $^2$P$_{3/2}$ & 3s & $^2$S$_{1/2}$&  17.322 &  0.17  & 0.064 & 0.25 & 0.77 & 0.082 & 0.12 & 0.28  & 1.10 & * \\ 


  40 &  3 &       2p$^{5}$ 3s 3p & $^4$P$_{5/2}$ &   3p & $^2$P$_{3/2}$ &  17.498  & 0.09  & 0.031 & 0.08 & 0.35 & 0.037 & 0.027 & 0.020 & 0.53 & * \\ 

  



  
  50 &  3 &       2p$^{5}$ 3s 3p & $^2$P$_{1/2}$ &   3p & $^2$P$_{3/2}$ &  16.855 & 0.15 & 0.075 & 0.16 & 0.98 &  0.083  & 3.2 & 4.3 & 1.31 & \\ 

  51 &  3 &       2p$^{5}$ 3s 3p & $^2$D$_{3/2}$ &   3p & $^2$P$_{3/2}$ &  16.895 & 0.09 & 0.031 & 0.09 & 0.52 &0.042 & 0.12  & 0.059 & 1.41 & \\ 

  73 &  4 &       2p$^{5}$ 3s 3d & $^4$F$_{5/2}$ &   3d & $^2$D$_{3/2}$ &  17.450 & 0.11 & 0.022 & 0.14 & 0.61 & 0.037 & 0.13 & 0.068 & 0.68 & * \\ 

  

  78 &  5 &       2p$^{5}$ 3s 3d & $^4$D$_{7/2}$ &   3d & $^2$D$_{5/2}$ &  17.411 & 0.16 & 0.023 & 0.21 & 0.72 &  0.037 & 0.18 & 0.098 & 0.70 & * \\ 

  79 &  5 &       2p$^{5}$ 3s 3d & $^2$F$_{5/2}$ &   3d & $^2$D$_{5/2}$ &  17.402 & 0.14 & 0.028 & 0.09 & 0.56 &  0.028 & 0.26 & 0.068 & 0.01 & * \\ 
  79 &  4 &       2p$^{5}$ 3s 3d & $^2$F$_{5/2}$ &   3d & $^2$D$_{3/2}$ &  17.393 & 0.12 & 0.023 & 0.07 & 0.56 &  0.023 &         &  & & \\ %

  80 &  4 &       2p$^{5}$ 3s 3d & $^2$P$_{1/2}$ &   3d & $^2$D$_{3/2}$ &  17.399 & 0.07 & 0.039 & 0.12 & 0.89 & 0.069 & 2.4 & 1.1 & 1.70 & \\ 
  80 &  1 &       2p$^{5}$ 3s 3d & $^2$P$_{1/2}$ &   3s & $^2$S$_{1/2}$ &  15.569 & 0.10 & 0.052 & 0.11 &      & 0.063 &  &  & & \\ %

  81 &  5 &       2p$^{5}$ 3s 3d & $^2$P$_{3/2}$ &   3d & $^2$D$_{5/2}$ &  17.368 & 0.08 & 0.038 & 0.13 & 0.43 &  0.078 & 0.40 & 0.30  & 1.86 & *  \\ 
  81 &  1 &       2p$^{5}$ 3s 3d & $^2$P$_{3/2}$ &   3s & $^2$S$_{1/2}$ &  15.536 & 0.62 & 0.28  & 0.54 &      & 0.31 &          &  & & *  IS \\ %
  
  82 &  1 &       2p$^{5}$ 3s 3d & $^4$D$_{1/2}$ &   3s & $^2$S$_{1/2}$ &  15.369 & 0.62 & 0.32 & 0.76 & 0.95  &  0.40 & 10 & 7.9 & 1.12 & \\ 
  
  83 &  1 &       2p$^{5}$ 3s 3d & $^4$D$_{3/2}$ &   3s & $^2$S$_{1/2}$ &  15.449 & 0.40 & 0.10 & 0.67 & 0.97  &  0.17 & 8.3 & 7.9  & 1.22 & \\ 

 
  84 &  5 &       2p$^{5}$ 3s 3d & $^2$F$_{7/2}$ &   3d & $^2$D$_{5/2}$ &  17.341 & 0.10 & 0.013 & 0.18 & 0.96 & 0.023 & 0.59 & 0.59 & 1.15 & \\ 
  
  85 &  4 &       2p$^{5}$ 3s 3d & $^4$F$_{3/2}$ &   3d & $^2$D$_{3/2}$ &  17.163 & 0.12 & 0.03 & 0.20 & 0.94 &  0.053 & 2.2 & 1.1 & 1.40 & \\ 
  85 &  1 &       2p$^{5}$ 3s 3d & $^4$F$_{3/2}$ &   3s & $^2$S$_{1/2}$ &  15.380 & 0.11 & 0.029& 0.07 &      &  0.019 &  &  & & \\ %
  
  

  88 &  5 &       2p$^{5}$ 3s 3d & $^2$F$_{7/2}$ &   3d & $^2$D$_{5/2}$ &  17.127 & 0.27 & 0.039 & 0.32 & 0.61  & 0.065 & 0.24 & 0.11  & 0.79  & * \\ 


  90 &  5 &       2p$^{5}$ 3s 3d & $^2$D$_{5/2}$ &   3d & $^2$D$_{5/2}$ &  17.130 & 0.12 & 0.031 & 0.13 & 0.23 & 0.097 & 0.099 & 0.043  & 0.81 & * \\ 
  90 &  4 &       2p$^{5}$ 3s 3d & $^2$D$_{5/2}$ &   3d & $^2$D$_{3/2}$ &  17.122 & 0.08  & 0.020 & 0.06 &  & 0.047 &  &  & & \\ %

  
  92 &  4 &    2p$^{5}$ 3p$^{2}$ & $^2$P$_{3/2}$ &   3d & $^2$D$_{3/2}$ &  16.904 & 0.07 & 0.017 & 0.12 & 1.0 & 0.030 & 34 & 35   & 0.84 &   \\ 
  92 &  1 &    2p$^{5}$ 3p$^{2}$ & $^2$P$_{3/2}$ &   3s & $^2$S$_{1/2}$ &  15.169 & 0.75 & 0.019 & 0.02 &     & 0.004 &   &  & & \\ %

  93 &  1 &       2p$^{5}$ 3s 3d & $^2$D$_{3/2}$ &   3s & $^2$S$_{1/2}$ &  15.314 & 0.64 & 1.3 & 4.3 & 0.69  & 1.6 & 0.19  & 3.6 & 0.56 & * IS \\ %

  94 &  1 &       2p$^{5}$ 3s 3d & $^2$P$_{1/2}$ &   3s & $^2$S$_{1/2}$ &  15.211 & 1.10 & 2.5 & 0.11 & 0.02  & 2.6 & 0.71 & 0.059  & 0.04 & * IS \\ 
  
  96 &  5 &       2p$^{5}$ 3s 3d & $^2$F$_{5/2}$ &   3d & $^2$D$_{5/2}$ &  17.014 & 0.07 & 0.013 & 0.19 & 0.90  & 0.036 & 0.33 & 0.41 & 0.99 & \\ 



 102 &  1 &       2p$^{5}$ 3s 3d & $^2$P$_{3/2}$ &   3s & $^2$S$_{1/2}$ &  15.163 & 3.6  & 1.1 & 3.4 & 0.84  & 1.0 & 5.3 & 5.6 & 0.60 & * IS \\ 

\noalign{\smallskip}
\hline
\end{tabular}
\begin{tablenotes}
    \item[] {The first columns give the upper $j$ and lower $i$ level number, the
  main configurations from the CHIANTI v.9 $n$=3 model and the CHIANTI v.9 wavelength (\AA)
of the transition. Column 5 gives the F2 value
  (only the strongest lines with values higher than 5x10$^{11}$ are shown.
  Note that single observed lines have typical values higher than 10$^{13}$).
  The following columns show the CHIANTI v.9 A-values and those
  with our improved $n=3$ model, and the total AI rates from the autionizing state.
  The final column gives R, the ratio between the AI rate as calculated with the
  same improved  $n=3$ model and with the $R$-matrix codes.  }
 \end{tablenotes}
\end{center}
\end{table*}

\begin{table*}
\begin{center}
   \caption{List of the  $n=3$ transitions formed by dielectronic capture with strongest
   intensity factor $F_2$ ($n$=6 model). 
\label{tab:table3} }
\footnotesize
\setlength\tabcolsep{4pt} 
\begin{tabular}{@{}rllllllllllllllllllllll@{}}
  \hline\noalign{\smallskip}
 $j$ & $i$&      C$_j$       & T$_j$               & C$_i$  & T$_i$  & $\lambda$ & $\lambda$(P) & $F_2$ & $Y$ &$F_2$(P)& $F_2$(C) &$A_{ji}$ & AI  & AI & AI  &  AI \\
     &     &                 &                           &        &        & (\AA)    &     (\AA)    & &   &      &       &   $n=6$    & $n=6$ & $n=3$ & B98 & N86 \\
\noalign{\smallskip}  \hline

82 &  1 &   2s$^{2}$ 2p$^{5}$ 3s 3d & $^4$D$_{1/2}$ & 3s & $^2$S$_{1/2}$    &  15.509 & 15.181? &  0.72 & 0.95& 1.5 ? & 1.1 ? &  0.38 & 7.1 & 7.9 & 7.9 & 8.6 & \\ %

 93 &  1 &   2s$^{2}$ 2p$^{5}$ 3s 3d & $^2$D[$^2$P]$_{3/2}$ & 3s & $^2$S$_{1/2}$& 15.269 &          &  3.9 & 0.67 &  &  &  1.5 & 3.1  & 3.6 & 2.0 & 5.0 &  IS \\ %

  99 &  1 &   2s$^{2}$ 2p$^{5}$ 3s 3d & $^2$P$_{3/2}$ & 3s & $^2$S$_{1/2}$ &  15.112 & 15.148 &  3.3 &0.84 &  2.4 & 3.8 &  0.99 & 5.5  & 5.6 & 3.3 & 4.2 & IS \\ %

140 &  3 &   2s$^{2}$ 2p$^{5}$ 3p 3d & $^4$P[$^2$D]$_{5/2}$ & 3p & $^2$P$_{3/2}$ &  15.432 & &  1.1 &0.98 &  &  &  0.18 & 7.4 & 5.7 &  5.2 & 4.5 &  \\ 

143 &  3 &   2s$^{2}$ 2p$^{5}$ 3p 3d & $^2$D$_{5/2}$ & 3p & $^2$P$_{3/2}$ &  15.399 & 15.029? & 0.58 &0.83  & 1.1? &  &  0.12 & 0.58 & 0.6 &  4.8e-3 & 0.013 & * \\ %

151 &  2 &   2s$^{2}$ 2p$^{5}$ 3p 3d & $^2$D$_{3/2}$ & 3p & $^2$P$_{1/2}$ &  15.255 & 15.221? & 7.3 &0.88  & 8.2? & 8.1? &  2.1 & 16  & 13 & 11 & 11 &  \\ %

152 &  3 &   2s$^{2}$ 2p$^{5}$ 3p 3d & $^2$D$_{5/2}$ & 3p & $^2$P$_{3/2}$ &  15.297 & 15.138? & 4.8 &0.85 & 3.0 &  &  0.95 & 5.3  & 4.8 & 3.9 & 4.2 & \\ %

153 &  2 &   2s$^{2}$ 2p$^{5}$ 3p 3d & $^2$P$_{1/2}$ & 3p & $^2$P$_{1/2}$ &  15.237  & 15.212 &  1.7 & 0.56 & 1.8 & 1.7 &  1.5 & 2.0 & 1.9 & 2.3 & 2.5 & *  \\ %

154 &  3 &  2s$^{2}$ 2p$^{5}$ 3p 3d & $^2$P$_{3/2}$ & 3p & $^2$P$_{3/2}$ &  15.245 & 15.146? & 0.29 & 0.04 & 2.8 ? & 2.4 ? &  1.8 & 0.077 &  0.025 & 8e-4 & 4.8e-3 & * \\ %

 156 &  3 &   2s$^{2}$ 2p$^{5}$ 3p 3d & $^2$P$_{1/2}$ & 3p & $^2$P$_{3/2}$ &  15.190  & &  1.5 &0.43 &  &  &  1.7  & 1.3 & 0.8 & 2.1 & 1.8 & * \\ %
 
 159 &  3 &   2s$^{2}$ 2p$^{5}$ 3p 3d & $^2$P[$^2$D]$_{3/2}$ & 3p & $^2$P$_{3/2}$ &  15.157 & &  1.9 &0.91 &  &  &  0.5 & 5.7  &  6.0 & 7.2 & 5.7 & \\ %

160 &  2 &   2s$^{2}$ 2p$^{5}$ 3p 3d & $^2$S$_{1/2}$ & 3p & $^2$P$_{1/2}$      &  15.066 & 15.069 &  1.1 & 0.76 & 1.4 & 1.4 &  0.7 & 3.1  & 2.9 & 4.2 & 4.0 & *\\ %

188 &  4 & 2s$^{2}$ 2p$^{5}$ 3d$^{2}$ & $^4$G$_{5/2}$ & 3d & $^2$D$_{3/2}$ &  15.525 & 15.496 ? &  1.8 & 0.97 & 1.4 & 1.4 &  0.31 & 14 & 14 & 16 & 14 &  \\ %
188 &  5 & 2s$^{2}$ 2p$^{5}$ 3d$^{2}$ & $^4$G$_{5/2}$ & 3d & $^2$D$_{5/2}$ &  15.536 & 15.464 ? &  0.53&      & 1.1  &  &  0.09 &  & & & & \\ %

191 &  5 & 2s$^{2}$ 2p$^{5}$ 3d$^{2}$ & $^2$G$_{7/2}$ & 3d & $^2$D$_{5/2}$ &  15.507 & 15.474 &  2.1 &0.97 & 2.2 & 1.7 &  0.27 & 8.0 & 7.7 & 9.2 & 8.5 & \\ %

192 &  5 & 2s$^{2}$ 2p$^{5}$ 3d$^{2}$ & $^4$F$_{5/2}$ & 3d & $^2$D$_{5/2}$ &  15.491 &      &  1.0 & 0.67 &  &  &  0.25 & 0.8 & 1.0 & 1.2 & 1.3 &  \\ %

193 &  1 &         2s 2p$^{6}$ 3s 3p & $^2$P$_{3/2}$ & 3s & $^2$S$_{1/2}$ &  14.008 & &  1.2 & 0.95  & &  &  0.31 & 6.7  & 6.0 & 7.2 & 8.2 & \\ %

196 &  4 & 2s$^{2}$ 2p$^{5}$ 3d$^{2}$ & $^4$D$_{1/2}$ & 3d & $^2$D$_{3/2}$      &  15.424  & 15.407 &  0.90 & 0.63 & 1.1 &  &  0.7 & 1.2 & 0.73 & 1.5 & 1.1 & * \\ %

198 &  5 & 2s$^{2}$ 2p$^{5}$ 3d$^{2}$ & $^2$P[$^4$S]$_{3/2}$ & 3d & $^2$D$_{5/2}$ &  15.385 & 15.376 & 0.85& 0.82 & 1.0 &  &  0.26 & 1.2 & 0.67 & 1.4 & 0.9 & * \\ %

199 &  5 & 2s$^{2}$ 2p$^{5}$ 3d$^{2}$ & $^2$F$_{7/2}$ & 3d & $^2$D$_{5/2}$ &  15.382 & 15.353 &  1.8 &0.93 & 2.1 & 2.0 &  0.24 & 3.4 & 4.2 & 4.5 & 5.2 &  \\ %

200 &  4 & 2s$^{2}$ 2p$^{5}$ 3d$^{2}$ & $^2$F[$^2$D]$_{5/2}$ & 3d & $^2$D$_{3/2}$ &  15.368 & 15.329 &  2.8 &0.88 & 7.9 & 8.8 &  0.5 & 5.1 & 3.7 & 6.3 & 2.7 & * \\ %
200 &  5 & 2s$^{2}$ 2p$^{5}$ 3d$^{2}$ & $^2$F$_{5/2}$ & 3d & $^2$D$_{5/2}$        &  15.379 &        &  1.1 &     &     &     &  0.2 & 5.1 & & & & \\ %

 201 &  4 & 2s$^{2}$ 2p$^{5}$ 3d$^{2}$ & $^2$F$_{5/2}$ & 3d & $^2$D$_{3/2}$ &  15.356 & &  5.5 & 0.97  &  &  &  0.9 & 32  & 34 & 6.4 & 39 &  \\ %

209 &  4 & 2s$^{2}$ 2p$^{5}$ 3d$^{2}$ & $^2$P$_{1/2}$ & 3d & $^2$D$_{3/2}$ &  15.240  & 15.227 &  1.4 & 0.52 &1.7  & 2.0 &  1.3& 1.5 & 1.6 & 1.3 & * \\ %

210 &  5 & 2s$^{2}$ 2p$^{5}$ 3d$^{2}$ & $^2$F$_{7/2}$ & 3d & $^2$D$_{5/2}$ &  15.243  & 15.205 &  10 & 0.96 & 11 & 11  &  1.3 & 35 & 36 & 42 & 39 & \\ %

211 &  5 & 2s$^{2}$ 2p$^{5}$ 3d$^{2}$ & $^2$D$_{5/2}$ & 3d & $^2$D$_{5/2}$ &  15.221 & 15.194 &  2.8 & 0.19  & 3.8 & 3.3&  2.5 & 0.59  & 0.61 & 0.66 & 0.62 &  \\ %

 212 &  4 & 2s$^{2}$ 2p$^{5}$ 3d$^{2}$ & $^2$D$_{3/2}$ & 3d & $^2$D$_{3/2}$ &  15.194 & 15.170 & 0.48 & 0.06 & 1.7 & 1.2 &  1.9 & 0.18 & 0.22 & 0.34 & 0.43 & * \\ %
 212 &  5 & 2s$^{2}$ 2p$^{5}$ 3d$^{2}$ & $^2$D$_{3/2}$ & 3d & $^2$D$_{5/2}$ &  15.205 & 15.178 & 0.18 &      & 1.2 &     &  0.7 &     & & & &  * \\ %

 213 &  4 & 2s$^{2}$ 2p$^{5}$ 3d$^{2}$ & $^2$P$_{3/2}$ & 3d & $^2$D$_{3/2}$ &  15.168 & 15.150 & 1.9 & 0.62 & 3.6 & 2.9 &  0.76 & 3.7 & 3.0 & 4.3 & 3.5 & \\ %
 213 &  5 & 2s$^{2}$ 2p$^{5}$ 3d$^{2}$ & $^2$P$_{3/2}$ & 3d & $^2$D$_{5/2}$ &  15.179 & 15.158 & 3.7 &      & 3.4 & 4.1 &  1.5  &     & &     &    & \\ %

 214 &  4 & 2s$^{2}$ 2p$^{5}$ 3d$^{2}$ & $^2$P$_{1/2}$ & 3d & $^2$D$_{3/2}$ &  15.077 & 15.097 &  1.0 & 0.81 & 1.5 & 1.2 &  0.59 & 2.6  & 2.2 &  3.3 & 2.6 & \\ %

\noalign{\smallskip}
\hline
\end{tabular}
\begin{tablenotes}
    \item[] {The first columns give the upper $j$ and lower $i$ level number, and the
  main configurations from the $n$=6 model. For the lower C$_i$ the 2s$^{2}$ 2p$^{6}$ is omitted.
The following two columns list the theoretical wavelengths (\AA) from the  $n$=6 model
and from \cite{phillips_etal:1997} (P). The following three columns list the F2 values from
the present  $n$=6 model, the Phillips et al.  and \cite{cornille_etal:1994} (C) ones, in units of 10$^{13}$.
 Only the strongest observable lines with F2 values larger than 10$^{13}$ are shown.
  The following columns show the  A-values and the total AI rates from the autoinizing state, also  in units of 10$^{13}$.
The AI $n=3$ is obtained with the $n=3$ set and KUTOO=1.  The B98 are the AI rates
calculated with Cowan's code by \cite{bruch_etal:1998} whilst the N86 are the YODA ones from \cite{nilsen:1989}.
An asterisc in the last column indicates  differences in the AI rates that can affect
the line intensity, as the ratio $Y$ differs from unity, with the exception of the transitions
where inner-shell (IS in the last column) is a dominant process.}
 \end{tablenotes}
\end{center}
\end{table*}

Table~\ref{tab:table3} lists all the main  $n=3$ transitions formed by dielectronic capture,
having an intensity factor $F_2$ larger than 10$^{13}$.
The results  are from the $n=6$ model. We compare our ab-initio
 wavelengths and $F_2$ values with those from \cite{phillips_etal:1997}, finding in some cases significant
 differences.
 We also compare our $F_2$ values with those from  \cite{cornille_etal:1994},
 finding a large scatter of values. We list a question mark when the level matching
 is unclear. 
 We believe that the main differences in the $F_2$ values are  due to the
 different AI rates used for the three calculations. 
We list in the last columns
 the AI rates  from two of our calculations, and those 
calculated with Cowan's code by \cite{bruch_etal:1998}
and with YODA by \cite{nilsen:1989}.
We see large discrepancies even for strong transitions. The most
important cases, highlighted with an asterisk in the last columnm,
are when the ratio $Y$ is far from unity. However,
with a few exceptions, the scatter of values is within 30\%.

\subsection{CE rates - DW vs. $R$-matrix}

\begin{figure}
\centerline{\epsfig{file=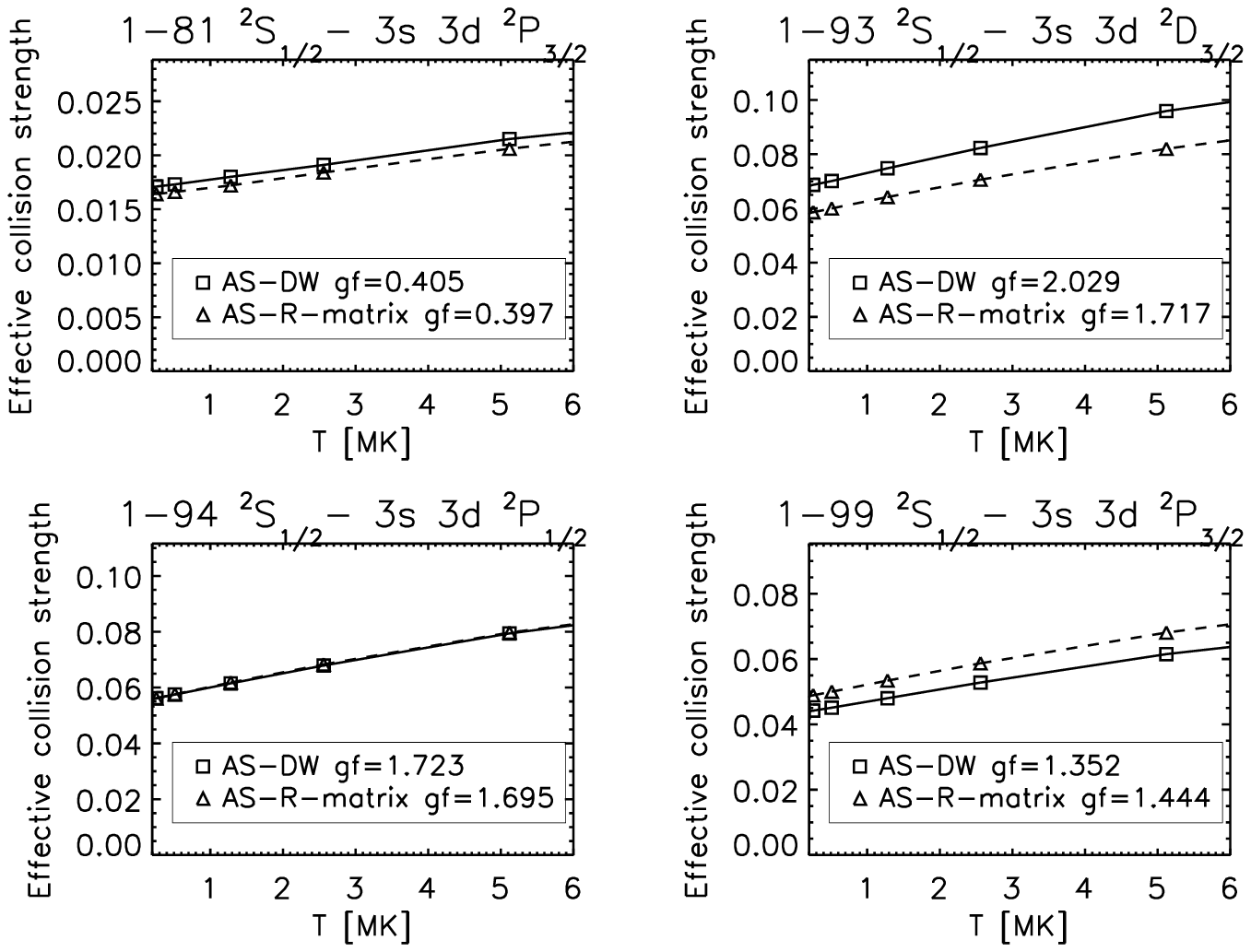, width=7.cm,angle=-90 }}
\caption{A comparison between the rate coefficients calculated with the DW
  approximation ($n=6$ model) and those with the $R$-matrix codes and the
 $n=3$ model calculated by Liang et al. }
\label{fig:ups}
\end{figure}

One important issue is whether the DW approach provides
accurate CE rates, compared to those obtained with the 
$R$-matrix codes. As the effect of the resonances is small for the AI states,
also considering the resonances is small for the AI states, one would expect
that the DW rates are accurate enough even for the $n=3$ states. 

We found that variations of the order of 20-30\% in the CE rates calculated with different
targets  are common, but are mostly related to
the variations in the oscillator strengths of each set of calculations,
as one would expect. 

We have compared the DW cross-sections and rate coefficients for key  transitions
formed by inner-shell excitation 
with those calculated by \cite{liang_etal:2008} using the
$R$-matrix codes and radiation damping, finding overall very good agreement.
 Figure~\ref{fig:ups} shows a comparison of rate coefficients for the
 strongest transitions, calculated with the $n=6$ model. The minor differences are related to
 the different $gf$ values, which are listed in the plot.

\section{Comparison with solar observations}

Clearly, in any laboratory or astrophysical spectra, the
satellite lines will always be blended with lines from the Ne-like iron
and other ions. As the Na-like abundance peaks around 2 MK in ionization equilibrium,
the best spectra would be those of 2 MK plasma. Unfortunately,
except for the very low resolution ones of MaGIXS, no solar spectra
of 2 MK plasma around 15~\AA\ exist.

Most of the spectra in this spectral region are of flares or
active regions where the peak temperature is at least 3 MK.
Chandra spectra of cool stars exist but do not have  resolution and
signal-to-noise comparable to solar spectra. 

The best astrophysical spectrum of the 15--17~\AA\ range, with the \ion{Fe}{xvii}
lines and their satellites  was taken by a 
Skylark sounding rocket on 30 Nov. 1971, with  Bragg crystal spectrometers
 built by the University of
Leicester (UK) under the supervision of Ken Pounds, pioneer of
X-ray astronomy. The spectrum was of a quiescent active region and
is high-resolution  as the instrumental FWHM was about 0.025~\AA.
Weak lines were measured, and the
wavelength and radiometric calibration was excellent. Details and a Table of wavelengths
and fluxes  are found in 
\cite[hereafter P75][]{parkinson:75}.
Some results from analyses of this spectrum were published by
\cite{delzanna:2011_fe_17} and \cite{delzanna_mason:2014},
where the P75 fluxes were converted to radiances.

The P75 spectra in the plots of the paper  appear  noisy, and do not clearly show all  the measured
line intensities listed in the published Table. On the other hand,
there are weak features in the spectra that are not listed in the Table.
To help the atomic data benchmarking procedure, we have created a reconstructed
P75 spectrum from the  list of the fluxes observed with the KAP crystal,
assuming a simple Gaussian  broadening with a FWHM=0.025~\AA.

The P75  wavelengths are so accurate, down to a few m\AA, that have been used
as reference  wavelengths for several
X-ray lines. Most  of the weaker P75 lines were unidentified. Several
turn out to be due to satellite lines from  \ion{Fe}{xvi},  as discussed below.

It is interesting to note that the poorly cited laboratory study by
\cite{cohen_feldman:1970} lists several lines which are within
a few  m\AA\  of the P75 ones. 
The spectrum was obtained with a 3m high-resolution spectrograph, from a 
low-inductance iron vacuum spark.
At the time most of the lines were unidentified, but the class of the line
listed by \cite{cohen_feldman:1970} and the wavelength coincidences with the
P75 suggests that several  of the satellite lines discussed below were also
observed in the vacuum spark.

\begin{figure}
\centerline{\epsfig{file=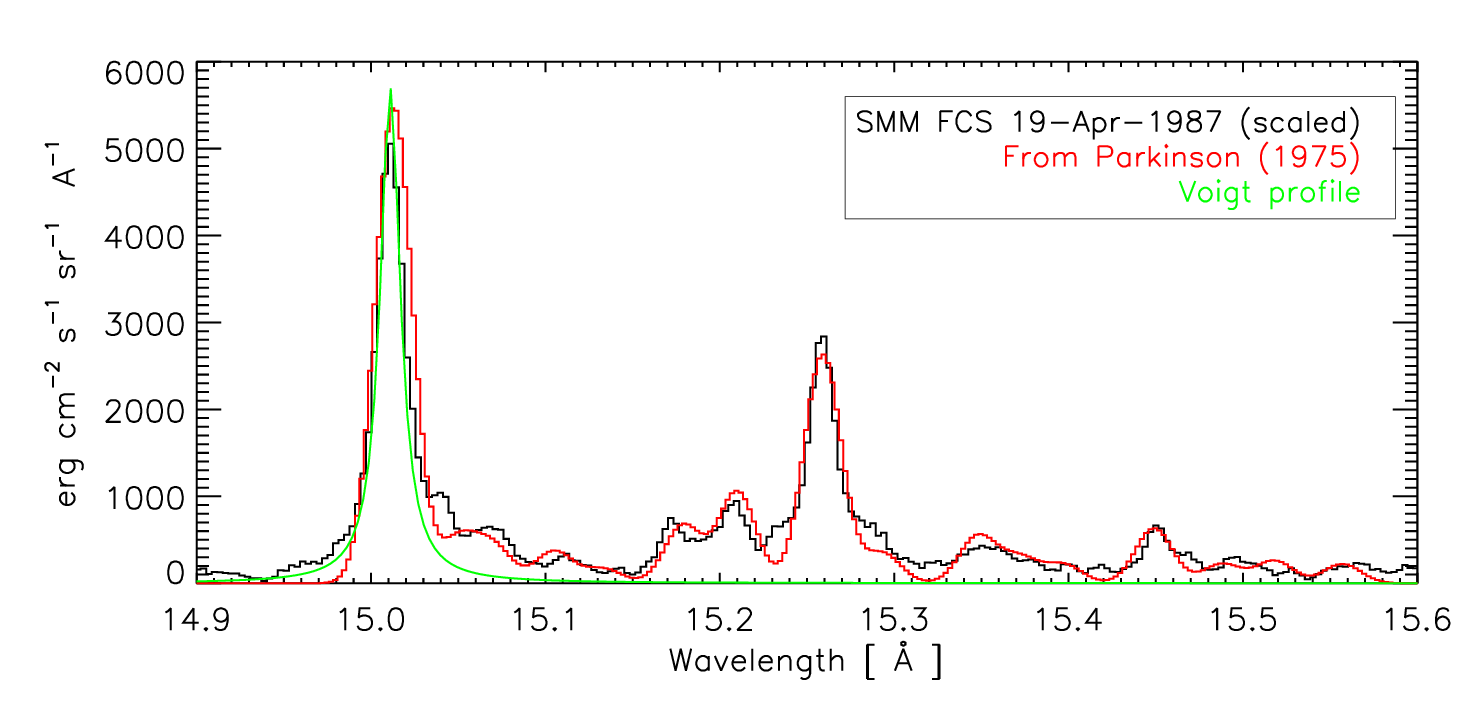, width=9.cm,angle=0 }}
\caption{A comparison between the averaged SMM/FCS spectrum of an active region
and that reconstructed from the fluxes tabulated by P75,
in the spectral region where the main \ion{Fe}{xvii} satellite lines are present.
The FCS instrumental Voigt profile of the \ion{Fe}{xvii} resonance
line is shown in green. }
\label{fig:b1}
\end{figure}

The  Solar Maximum Mission (SMM) Flat Crystal Spectrometer (FCS) also
produced spectra of quiescent active regions, but with too short exposures
and low signal. The instrument was always pointed at the brightest
parts, where the hot core loops typically have 3-4 MK. 
We have searched the entire SMM FCS database for suitable observations of the satellite
lines, but encountered the following problems. First, the count rates of a single bin
of each crystal scan are very low, hence to increase the signal to noise several
scans need to be averaged. Second, significant variability in the strong lines is often
present. In our previous analysis of FCS data \citep{delzanna_mason:2014},
we focused on very stable and quiescent
scans, where variability is reduced. However, the signal in the satellite lines is too low
in those spectra.

We have also analysed spectra during, or after, large flares. In those cases the signal
in the lines is generally higher, but that in the satellite lines is lower.
Additionally, hotter lines from e.g. \ion{Fe}{xviii} and \ion{Fe}{xix}
appear and further complicate  the analysis of the spectra.

We found that  the best case is a series of five scans taken on 1987-Apr-19 between
15:54 and 16:42 UT. Some variability was present, and three spectra were scaled
by small amounts (10, 20, and 30\%)  before averaging, to compensate for the
variability.
The spectrum  was then smoothed,  converted to calibrated units, and increased by a
factor of 2.7 to match approximately the reconstructed AR spectra of  P75.
The two spectra, plotted in Figure~\ref{fig:b1},  show a remarkable
similarity. The P75 spectrum was clearly of much higher quality, as even the
weakest satellite lines were measurable (with about 50 total counts).

The main difference between the observed and reconstructed P75 spectra are the
broad features surrounding the  \ion{Fe}{xvii} 15.01 and 15.26~\AA\ 3C and 3D lines.
The FCS  instrumental line profile is well approximated with a Voigt function.
We have taken the profile of the  \ion{Fe}{xvii}  16.75~\AA\ line and fitted
with a Voigt profile, which is also plotted (rescaled by the peak intensity) 
in  Figure~\ref{fig:b1}, to show that in fact there is significant signal
in the wings of the resonance line at 15.01~\AA, especially in the
red wing, with two broad features at 15.04 and 15.07~\AA,
which were listed at 15.05 and 15.07~\AA\ by P75.
The main difference is that the broad feature at 15.04~\AA\ is much stronger
than the other one, in the FCS spectra. This results from the slightly better
spectral resolution of the FCS instrument.

Something similar is present in the red wing of the 15.26~\AA\ line,
where P75 list a feature at 15.293~\AA, although  the FCS spectrum indicates a
broader and brighter feature. The FCS spectrum  also shows a broad feature in the blue
wing of the 3D 15.26~\AA\ line. 
Inspection of the spectra in Fig.1b of P75 also indicates
broadenings around the two   \ion{Fe}{xvii} lines.

\begin{figure}
\centerline{\epsfig{file=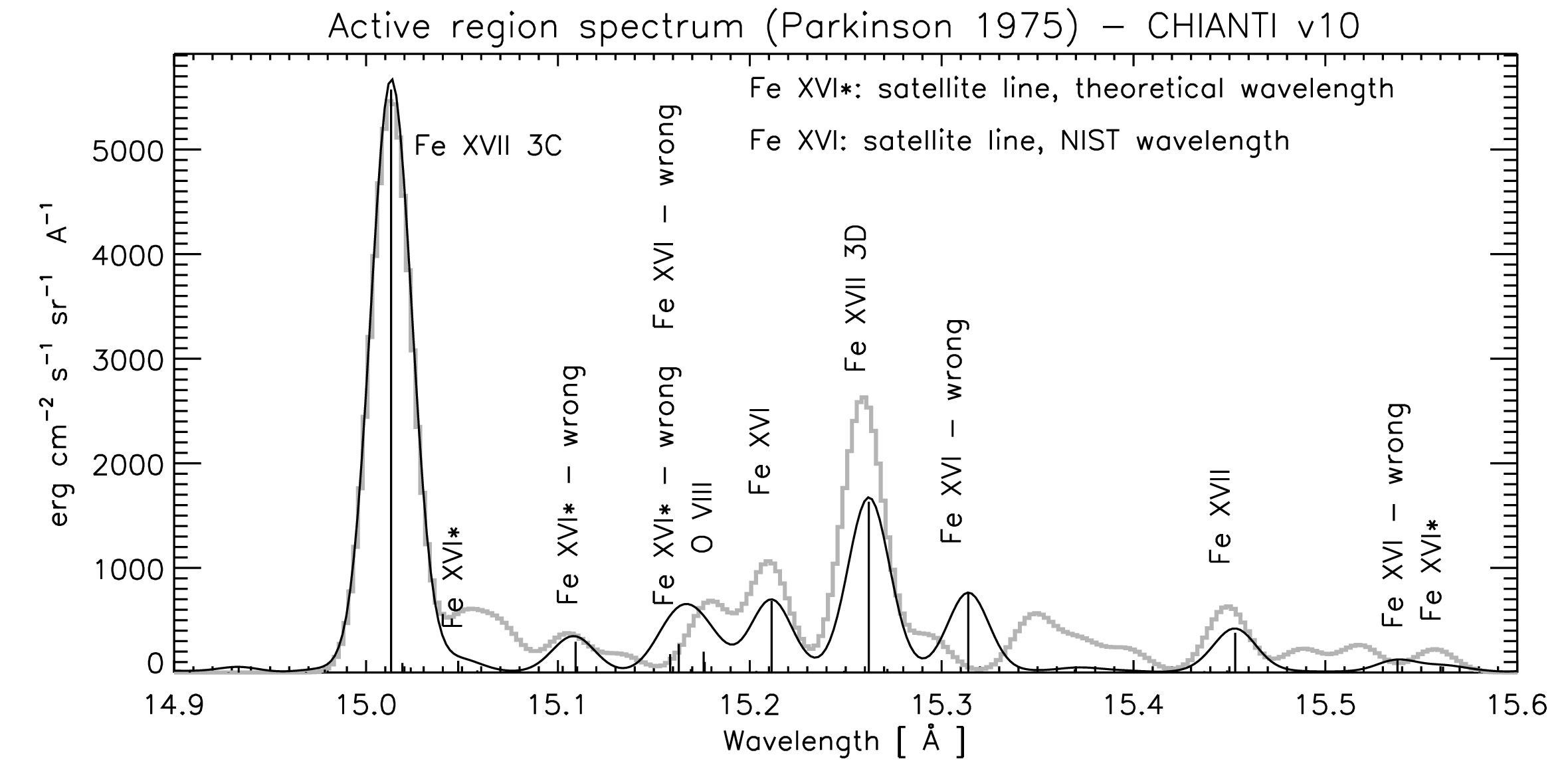, width=9.cm,angle=0 }}
\centerline{\epsfig{file=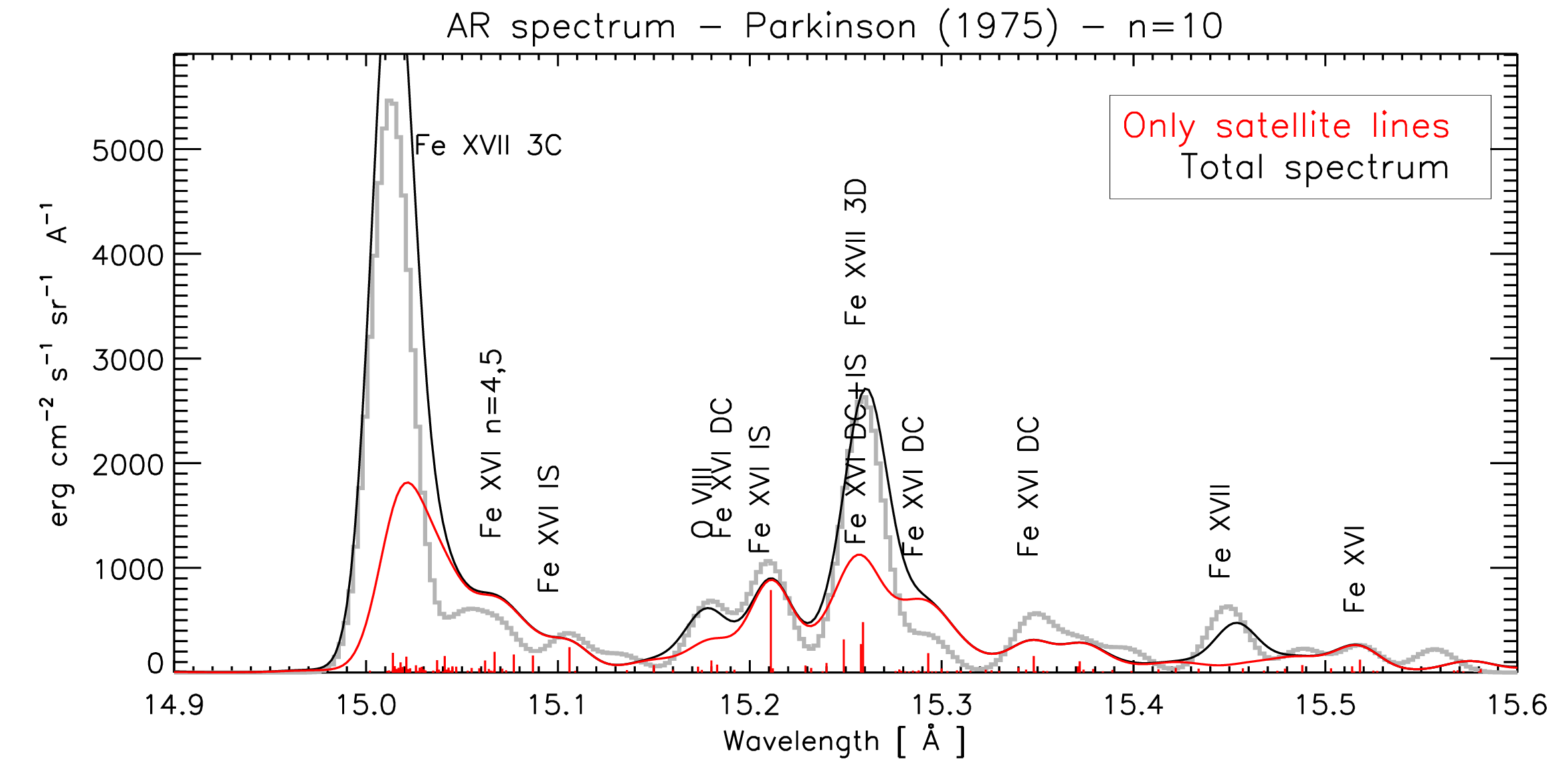, width=9.cm,angle=0 }}
\centerline{\epsfig{file=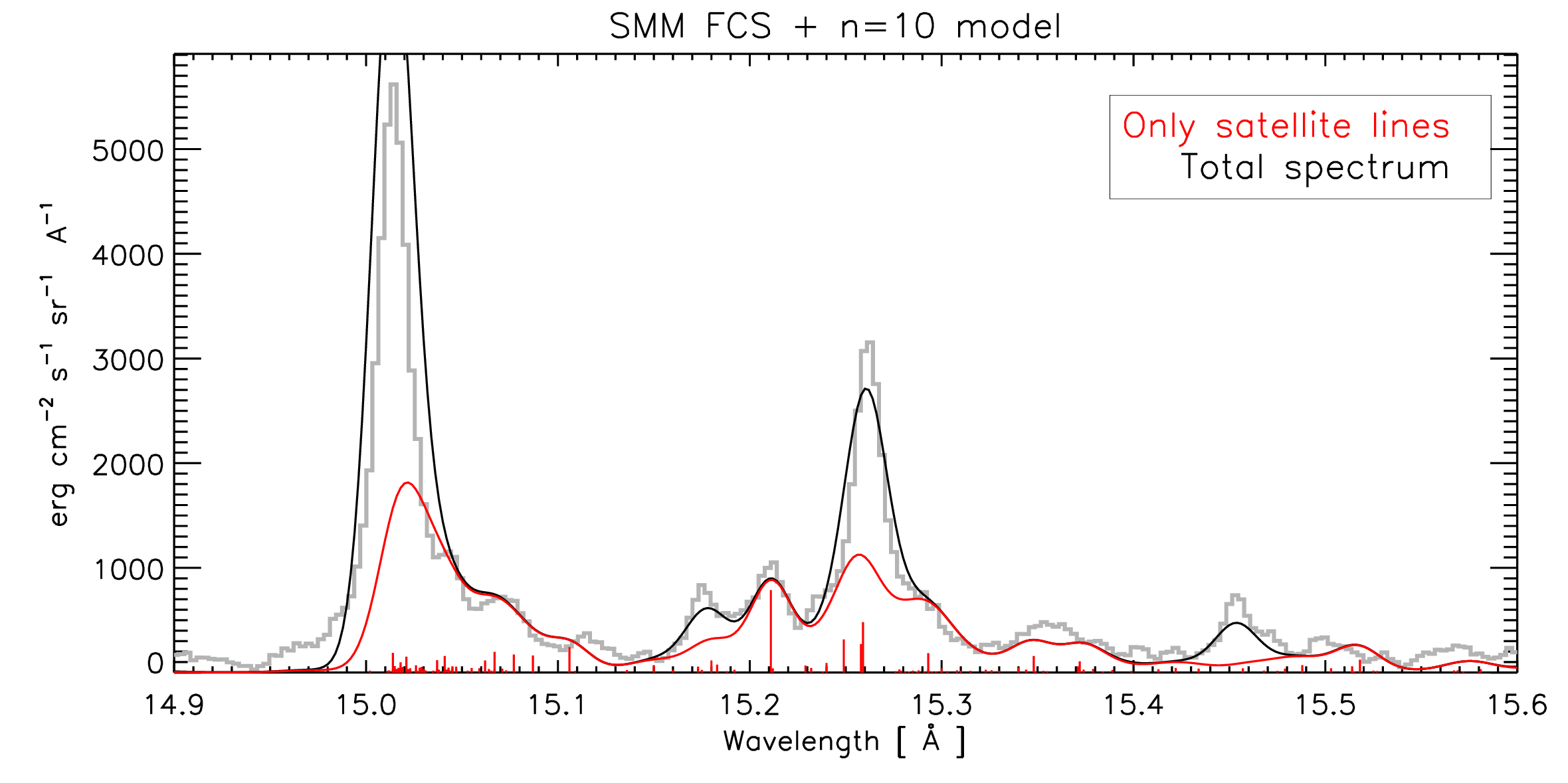, width=9.cm,angle=0 }}
\caption{A comparison between the reconstructed P75 spectrum (grey thick line) and an
averaged SMM/FCS spectrum of an active region with the 
CHIANTI v.10 data (top plot)  and the present $n=10$ model
in the 15-15.6~\AA\ region. The CHIANTI (NIST) wavelengths of the
main lines are incorrect, with one exception. 
A few satellite lines mainly formed by inner-shell (IS)
excitation or dielectronic capture (DC) are indicated. }
\label{fig:sp15}
\end{figure}

\begin{figure}
\centerline{\epsfig{file=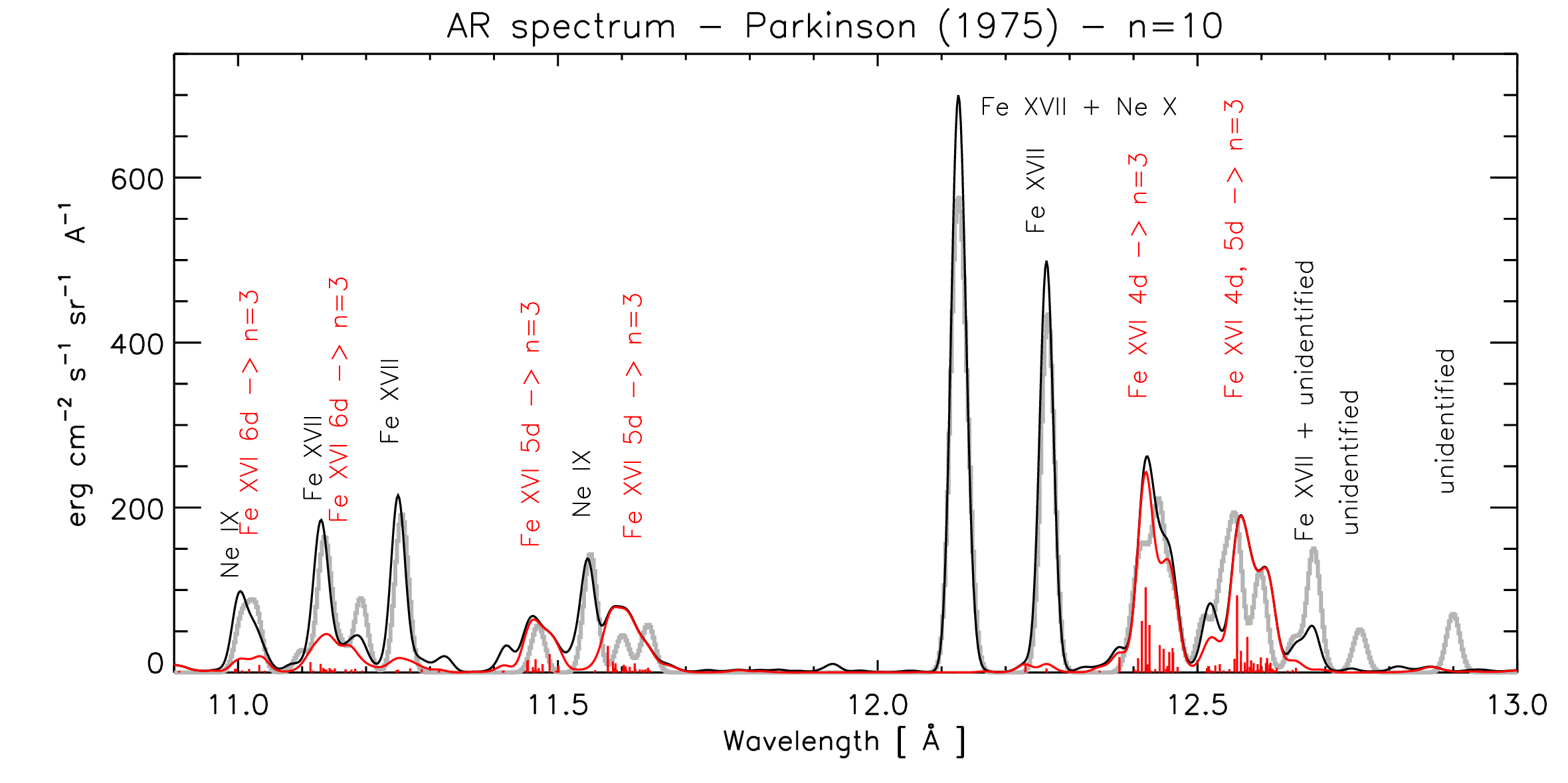, width=9.cm,angle=0 }}
\caption{A comparison between the reconstructed P75 spectrum (grey thick line)
and the present $n=10$ model, indicating that several previously unidentified lines are due to
Fe XVI. The red spectrum is the contribution from the satellite lines. 
}
\label{fig:sp}
\end{figure}

\begin{table*}
\begin{center}
   \caption{List of the strongest satellite lines from AI states of Fe XVI. 
\label{tab:p75} }
\begin{tabular}{@{}rllllllllllllllllllllll@{}}
  \hline\noalign{\smallskip}

Present     &  B79   &  P75 & CF70  & G09    & Transition (lower-upper) & $I_o$ & $I_p$  & levels  \\ 
\noalign{\smallskip} \hline

12.42   & -  & 12.414 & 12.411? & -  & 3s $^2$S$_{1/2}$ - 2s$^{2}$  2p$^{5}$ 3s 4d $^2$P$_{1/2}$ & 3.4 &  -,1.5,1.7,1.6,1.7 & 1--(336)380 \\
12.42   & -  & 12.414 & 12.411? & -  & 3s $^2$S$_{1/2}$ - 2s$^{2}$  2p$^{5}$ 3s 4d $^2$P$_{3/2}$ &     &  -,1.2,? ,2.7,2.8  & 1--(334)378 \\ 

12.43  & -   & -       & 12.427?  &  -   & 3p $^2$P$_{3/2}$ - 2s$^{2}$ 2p$^{5}$ 3p 4d $^2$D$_{5/2}$ &     & -,2.4,1.6 ,1.5,1.5  & 3--(522)566 \\ 

12.56  &  -  & 12.560  & -  &  - &  3s $^2$S$_{1/2}$ - 2s$^{2}$ 2p$^{5}$ 3s 4d $^4$D$_{3/2}$       & 4.7 &  -,3.0,2.7,2.5, 2.5 & 1--(298)342 \\

%

14.00  & -  & 14.037  & 14.020?  & 14.03?  & 3s $^2$S$_{1/2}$ -  2s 2p$^{6}$ 3s 3p $^2$P$_{3/2}$ & 4.5 &  -,0.8,1.3,1.2,1.3  & 1--(193)237 \\

\noalign{\smallskip}
15.11  & 15.163p & 15.105  & 15.110 & 15.111  & 3s $^2$S$_{1/2}$ -  2s$^2$ 2p$^{5}$ 3s 3d $^2$P$_{3/2}$ & 9.8 & 7.2,7.4,6.6,6.7,6.5  & 1--(99)143 \\ 


15.17 & -     &   ?      & 15.158? &  -  &  3d $^2$D$_{3/2}$ -  2s$^{2}$ 2p$^{5}$ 3d$^{2}$ $^2$P$_{3/2}$ &  & -,2.2,1.7,1.6,1.5  & 4--(213)257 \\ 

15.18 & -     & 15.179 & 15.173? & -   &  3d $^2$D$_{5/2}$ -  2s$^{2}$ 2p$^{5}$ 3d$^{2}$ $^2$P$_{3/2}$ & 18 &  -,4.1,3.1,3.2,3.0 & 5--(213)257 \\ 
15.18  & -    & 15.179  &  -   & -     & 3p $^2$P$_{3/2}$ - 2s$^{2}$ 2p$^{5}$ 3p 3d $^2$P$_{1/2}$            &   &  3.0,2.5,2.4,2.1,2.0  & 3--(156)200 \\

\noalign{\smallskip}
15.21 & 15.211p & 15.210  & 15.222? & 15.210 & 3s $^2$S$_{1/2}$ - 2s$^{2}$ 2p$^{5}$ 3s 3d $^2$P$_{1/2}$ & 28 & 18,20.9,21.7,21,21  & 1--(94)138 \\

\noalign{\smallskip}
15.24  &   -    & 15.259(bl 3D)  & -     &    -    & 3d $^2$D$_{5/2}$ -  2s$^{2}$ 2p$^{5}$ 3d$^{2}$ $^2$D$_{5/2}$ & & -,3.4,2.4,2.6,2.4 & 5--(211)255 \\

15.25  &    -   &  15.259(bl 3D)  & 15.237?&   -    & 3d $^2$D$_{5/2}$ -  2s$^{2}$ 2p$^{5}$ 3d$^{2}$ $^2$F$_{7/2}$ & 70 & -,12.0,8.9,8.9,8.5 & 5--(210)254 \\


15.26   & -     & 15.259(bl 3D)& 15.261 &        & 3p $^2$P$_{1/2}$ - 2s$^{2}$ 2p$^{5}$ 3p 3d $^2$D$_{3/2}$ &  &   8.0,10.6,7.8,7.8,7.3 & 2--(151)195 \\ 
15.26  &15.314p & 15.259(bl 3D)& 15.261 & 15.261 & 3s $^2$S$_{1/2}$ - 2s$^{2}$ 2p$^{5}$ 3s 3d $^2$D$_{3/2}$ &  &  20,13.8,11.9,13.7,13 & 1--(93)137 \\ 

\noalign{\smallskip}
15.29  &  -   & 15.293(bl) & 15.288? &  -      &  3p $^2$P$_{3/2}$ -  2s$^{2}$ 2p$^{5}$ 3p 3d $^2$D$_{5/2}$ & 9.3 &  4.8,6.9,5.0,5.2,4.9 & 3--(152)196 \\

15.35 & -  & 15.348  & 15.341 & - & 3d $^2$D$_{3/2}$ -  2s$^{2}$ 2p$^{5}$ 3d$^{2}$ $^2$F$_{5/2}$ & 14 &  -,3.6,2.7, 4.4, 4.2  & 4--(201)245 \\


15.37 & -  & 15.348  & 15.341  & - & 3d $^2$D$_{3/2}$ -   2s$^{2}$ 2p$^{5}$ 3d$^{2}$ $^2$F$_{5/2}$ & &  -,5.8,4.3,3.0,2.8 & 4--(200)244 \\ 


15.49  & -  &  15.488 & 15.490? & - & 3d $^2$D$_{5/2}$ -  2s$^{2}$ 2p$^{5}$ 3d$^{2}$ $^2$G$_{7/2}$ & 5.9 &  -,1.9,2.4,2.0,1.9  & 5--(191)235 \\



15.52  & 15.538p & 15.518  & - & 15.516 & 3s $^2$S$_{1/2}$ - 2s$^{2}$ 2p$^{5}$ 3s 3d $^2$P$_{3/2}$ & 6.9 &  3.1,5.3,5.3,3.3,3.3  & 1--(81)125 \\ 
\noalign{\smallskip}



\hline
\end{tabular}
\begin{tablenotes}
    \item[] {The first column gives the present wavelengths. The following three columns indicate 
    the possible wavelength matches from
    \cite{burkhalter_etal:1979} [B79, p for predicted] and from the lists of  the unidentified lines in
    \cite{parkinson:75} [P75] and \cite{cohen_feldman:1970} [CF70].
The next column lists
    the observed wavelengths from \cite{graf_etal:2009} [G09].  $I_o$ indicates the
    radiance in ergs cm$^{-2}$ sr$^{-1}$ s$^{-1}$ from P75, as described in \cite{delzanna:2011_fe_17},
    while $I_p$ list the predicted values (the first from the CHIANTI model, then those obtained
    from the $n=4,5,6,10$ models. The last column indicates the indices of the transition,
    relative to the $n=6$ (in brackets) and $n=10$ models. 
    }
 \end{tablenotes}
\end{center}
\end{table*}

\begin{table}
\begin{center}
   \caption{List of newly identified satellite lines, Parkinson's KAP crystal.
\label{tab:p75b} }
\begin{tabular}{@{}rllllllllllllllllllllll@{}}
  \hline\noalign{\smallskip}
$\lambda_{\rm obs}$(\AA) & Int & P75 & upper \\
\noalign{\smallskip} \hline
10.742 & 1.3  & Ne IX & bl $n$=7,8  \\
  11.006 & 1.7 & Na X & bl $n$=6  \\
  11.027 & 2.0 & Ne IX & bl $n$=6  \\
  11.099 & 0.7 &  Na X  & bl $n$=6 \\
  11.135 & 4.3 & Fe XVII &  bl $n$=5,6 \\
  11.160 & 0.9 & unid. &  $n$=6  \\
  11.192 & 2.4 & Na X  &  $n$=4  \\

11.469 & 1.5  & unid. &  $n$=5  \\
11.601 & 1.2  &  unid. & $n$=5,6  \\
11.641 & 1.5   &  unid. & $n$=5  \\

 12.399 & 1.2   &  unid. & $n$=4  \\
  12.414 & 3.4  &  unid. & $n$=4  \\
  12.439 & 5.2  &  unid. & $n$=4  \\
  12.463 & 2.4  &  unid. & $n$=4  \\
  12.510 & 1.8  &  unid. & $n$=4,5  \\
  12.539 & 2.8 &  unid. & $n$=4,5  \\
  12.560 & 4.7 &  unid. & $n$=4,5  \\
  12.598 & 3.3  &  unid. & $n$=4,5  \\
  12.651 & 1.1 & Ni XIX & bl $n$=4,5  \\
 13.868 & 2.1 & unid. & $n$=4,5   \\
 13.899 & 5.8 & Fe XVII & bl $n$=3,4,5   \\
  14.037 & 4.5 & Ni XIX & bl $n$=3  \\
  14.081 & 3.4 & Ni XIX & bl $n$=3  \\

  \noalign{\smallskip}
  17.16 & 8.9  & unid. & $n$=3  \\
  17.21 & 12.  & unid. & $n$=3  \\
  17.40 & 7.9  & unid. & $n$=3  \\
  17.51 & 5.1  & unid. & $n$=3  \\
  17.77 & 5.4  & O VII & bl $n$=3  \\
\hline
\end{tabular}
\begin{tablenotes}
\item[] {The columns give the observed wavelengths,
  the  radiance in ergs cm$^{-2}$ sr$^{-1}$ s$^{-1}$, P75
  identification (unid. for unidentified lines)
  and the upper states of the main satellite lines. bl indicates
blending with satellite lines.}
 \end{tablenotes}
\end{center}
\end{table}

We have slightly improved the previous emission measure analyses of the P75 data 
described in \cite{delzanna:2011_fe_17} and \cite{delzanna_mason:2014}, 
and calculated a predicted spectrum. 
We used CHIANTI v.10 \citep{chianti_v10} data, except for the satellite lines. 
Table~\ref{tab:p75} lists a selection of the strongest lines within the P75
spectral range. We list our ab-initio AS wavelength, and the
wavelengths of the lines we identify in the  \cite{parkinson:75} and  \cite{cohen_feldman:1970}
spectra. Whenever possible, we are also listing the observed wavelengths from \cite{graf_etal:2009}.
In the last columns  we provide the radiances, as measured by P75 and as calculated with the
various models.

Clearly, all the satellite lines are blended to some degree, so the
comparisons with the spectra, shown in Figure~\ref{fig:sp15}, are more instructive.
We have adjusted the energies
of a few main states (cf. the energy table) to produce the spectra.

The top plot in Figure~\ref{fig:sp15} clearly indicates that the 
 CHIANTI (NIST) wavelengths of the
main lines are incorrect, with one exception; and that, even for the
active region spectra, there is significant missing flux due to the
satellite lines.  The new $n=10$  model increases the flux significantly, and brings
the predicted intensity of the 3D line in to good agreement with observations.

Figure~\ref{fig:sp15_n6} in the Appendix shows the results from the
$n=6$  model, to show the effects of the additional $n=7$ -- $10$ configurations
in blending the 3C resonance line. 

We have calculated the total flux of the Fe XVII and the satellite lines in the 15--15.7~\AA\
range at 2 MK and found an increase of a factor of 1.93 with the $n=10$ model
(a factor of 1.8 with the $n=6$ one), relative to the
total flux of the current  CHIANTI model.
Indeed, at such low temperatures, the satellite lines dominate this
spectral range.

We discuss below a few details about the main lines, noting that
it is nearly impossible to list all previous identifications, whether
correct or incorrect.
The labelling is  from the $n=6$ model. 
We are relatively confident about our identifications,
but ultimately new high-resolution laboratory and solar spectra 
will be needed to confirm the present work.

\subsection{The 15--16~\AA\ region}

The strongest line is the decay to the ground state of the
3s3d~$^2$P$_{1/2}$ (level No. 94), mainly formed by inner-shell excitation.
The energy of the upper state was estimated by  B76 from the decay to the
3d $^2$D$_{5/2}$, observed at 16.952~\AA. The AS ab-initio wavelength
is very close (15.21) to the value estimated by B76 (15.211) and to the solar
and lab measurements by P75 and G09 (15.210). The predicted radiance is 22, close
to the observed one (28.1).

The second strongest line is the decay to the ground state of the
2s$^{2}$ 2p$^{5}$ 3s 3d $^2$D$_{3/2}$ (level No 93), also formed by inner-shell excitation.
The  B76 predicted wavelength is 15.314~\AA, from
 an energy of 6530000., obtained  from the decay to the 3d $^2$D$_{5/2}$ at 17.087~\AA.
 This identification is clearly incorrect for various reasons. First,
 there is no solar line at 15.314~\AA. Second, the predicted energy is very
 far from our  ab-initio value. 
 Given the predicted intensity and wavelength, this line must be blending
 the strong \ion{Fe}{xvii} 3D observed by P75 at 15.259~\AA.
 The same conclusion was obtained by G09, and earlier by \cite{brown_etal:2001}. 

Our model predicts a nearby strong line (2--151).  We assume it is
also blended with the \ion{Fe}{xvii} 3D line, which provides an excellent
comparison between predicted and observed spectra.
With this identification, we obtain an energy of 
6830703 cm$^{-1}$, close to the value of a state with the same J value and parity
given by D13 at 6831282 cm$^{-1}$. This line is not listed by G09.
What seems to be the same transition (Na5) was predicted by M05 to be at
a very similar wavelength, 
15.255~\AA, but was actually identified with a line at 15.276~\AA.
As there is no strong solar line at 15.276~\AA, the M05 identification seems incorrect.

The third strongest line, mostly formed by inner-shell excitation,
is the decay to the ground state from the 2p$^{5}$ 3s 3d $^2$P$_{3/2}$ (level No 99),
with an AS  predicted wavelength of 15.11~\AA.
The energy of the upper state was estimated by  B76 to be
6595000  from the decay to the  3d $^2$D$_{3/2}$, observed at 16.890~\AA.
This predicts the decay to the ground state  at 15.163~\AA, which is not
observed. Again, this was an incorrect identification by B76.
We identify, on the basis of wavelength and intensity, with the
P75 solar line at 15.105~\AA, observed in the laboratory by CF70 and G09
at 15.11~\AA.

There are several other cases where the B76 / NIST identifications are
incorrect. One clear case, where lines are not too blended, is the
decay from the 3s 3d $^2$P$_{3/2}$ (level No. 81). The AS predicted wavelength
is 15.54~\AA. It could either be the solar line at 15.518 or that one at 15.557~\AA,
while B76 predicted, on an incorrect identification of the
decay to the  3d $^2$D$_{5/2}$, a wavelength of 15.538~\AA, where a solar line
is not observed.
We favour the first option, as the wavelength agrees with the \cite{diaz_etal:2013}
calculations, and its intensity is well matched. 
G09 has a predicted wavelength of 15.533~\AA\ but
gives an observed wavelength of 15.516~\AA, despite the fact that
there is no line clearly visible  at that wavelength in their
spectrum of Fig.~2.

Our model predicts many strong lines from the 2p$^{5}$ 3d$^{2}$
configuration, which was not included by  \cite{liang_etal:2008}.
Such states were not considered by G09, as the lines are
formed by DC.  Surprisingly, such states were  also not considered  by B76.
M05 list only 
two transitions, from $J=5/2$ to 3d $^2$D$_{3/2}$ (Na2b, 15.360~\AA),
and from $J=3/2$ to 3d $^2$D$_{3/2}$, at a predicted wavelength of 15.200~\AA.
In our model, we actually  have two strong transitions from
two nearby $J=5/2$ states  to the  3d $^2$D$_{3/2}$, at
predicted wavelengths of 15.34 and 15.36~\AA. 
They are likely  blended to form the strong solar line at
15.348~\AA.  The model spectrum agrees very well with the
P75 nand FCS spectra. On the other hand, transitions  from $J=3/2$ to 3d $^2$D$_{3/2}$
are relatively weak in our model.

G09 found a blue wing around 15.19~\AA\ and identified it with
the transition 3--155 in their model, with a relative intensity of 0.18.
The present model instead predicts a very weak relative intensity of 0.03. 
On the other hand, our model predicts two strong decays from the 2s$^{2}$ 2p$^{5}$ 3d$^{2}$
( 5--211 and 5--210) which are most likely blending the 15.21 and 15.26~\AA\ lines.
We obtain a good agreement between predicted and observed spectra with these identifications.
One of the two lines, the 5--211, was instead identified by M05 with a line in the
Hercules spectra at 15.237~\AA from a predicted wavelength of 15.225,
although their  spectra also have a strong line at 15.213~\AA.
The M05 identification is inconsistent with the present model, as there is no strong solar
line at 15.237~\AA.

We also identify a  strong transition (3--152), with predicted wavelength of
15.30~\AA\  with the solar one observed
by P75 at 15.293~\AA\ (plus a blend of $n$=5 lines), considering that  intensities and wavelengths match.
 This line is not listed by G09.
What seems to be the same transition (Na3) was predicted by M05 to be at
a very similar wavelength, 
15.290~\AA, but was actually identified with a line at 15.304~\AA.

\subsection{The satellites blending the 3C line at 15~\AA. }

We note that the large number of transitions from $n=4,5$ states
provide ab-initio wavelengths and intensities in broad
agreement with the FCS observation.
Several transitions from $n=4,5$ states are either blending
the resonance line, or  are scattered across many different spectral
ranges.  

The $n=4$ satellites to  the  3C line  are mostly resolvable, being in the
red wing, but the  others are not. 
As previously mentioned, we have carried out a large-scale $n=10$
calculation to improve the estimate of all the satellites
of the 3C line. Most of the flux is due to transitions from
$n=4,5$ states, but as \cite{beiersdorfer_etal:2011} pointed out,
some contribution from higher states is also present.

If one considers only the 2p$^5$ 3d $nl$ satellites
within the 15.0--15.06~\AA, which contribute about 75\% of the
total flux in this band with the  $n=10$ model, the  $n=4,5,6$ states contribute
most, with 16, 33 and 15\% respectively. The
$n=7,8,9,10$ states  contribute progressively less:
9.5, 8.3, 5.3, and 3.9\%.
Therefore, any missing flux due to even higher states
would likely be at most a few percent. 

Finally, to assess the possible missing flux, we have carried out a
configuration-averaged complete calculation including configurations up to $n=30$.
The wavelengths of the transitions are not very accurate, and
if one considers decays from individual  configurations
some discrepancies are found with the totals
from the j-resolved  $n=10$ model, hence comparisons between the calculations are not simple. 
If one considers the main transitions from the  2p$^5$ 3d $nl$ states,
the $n=30$ model indicates that those from the $n=4-10$ states
contribute 91\%, those from the $n=11-20$ states 6\% and those
from the  $n=21-30$ states 3\%.
However, the total flux within the 3C line from the $n=30$
model is less than 2\% larger than what we calculated with the  $n=10$ model.
Therefore, we conclude that any missing flux within the $n=10$ model
would amount only to a few percent, in broad agreement with the
\cite{beiersdorfer_etal:2011} FAC result.

\subsection{The 10--15~\AA\ region}

As shown in  Figure~\ref{fig:sp}, our model predicts many
transitions blending known transitions or explaining previously
unidentified lines. Table~\ref{tab:p75b} provides a short summary.
There are many transitions from $n=6$ states that are blending known
lines from Ne IX and Fe XVII between 11.0 and 11.3~\AA.
Within the 11.45 and 11.50~\AA\ and
 11.60 and 11.65~\AA\ regions, our model predicts
several decays from  2s$^2$~2p$^{5}$~3$l$~5d states. 
Parkinson's spectrum has indeed unidentified transitions
at 11.469, 11.601, and 11.641 with intensities similar
to the predicted ones.

Within the 12.3 and 12.7~\AA\ region, there are several relatively
strong transitions from 2s$^{2}$~2p$^{5}$~3$l$~4$l'$ and some $n=5$ states.
Parkinson's spectrum has many  unidentified transitions 
in this wavelength range. There is an excellent match
between our ab-initio wavelengths and calculated intensities and the
observed spectrum. 


Several decays from 1s$^{2}$~2s~2p$^{6}$~3p 5f and 5g states
are predicted between 13.8 and 13.9~\AA\ and are
 blending the Fe XVII 13.899~\AA\ line, which is
significantly under-predicted, and form the unidentified line at  13.868~\AA.

Several transitions of the type
1s$^{2}$~2s~2p$^{6}$~3$l$~3$l''$ are predicted around 14.0~\AA, are 
blending  Ni XIX lines around 14.04~\AA.

\subsection{The 17--18~\AA\ region}

The situation for the weaker lines above 17~\AA\ is rather unclear,
as the calculations are more uncertain, and the sensitivity of the
solar instruments was lower, hence only a few of the stronger lines were observed.
G09 list various identifications for lines formed
by inner-shell excitation, but agreement with the observed
spectrum is poor. P75 only lists four unidentified lines,
at 17.151, 17.199, 17.389 (blended with \ion{O}{vii}), and 17.501~\AA.
However, all the lines in Parkinson's spectrum above 16~\AA\
have an incorrect wavelength, being lower by 0.01~\AA.
We have applied such correction to the spectra, see
Table~\ref{tab:p75b}. 

G09 lists several possible lines in the 17.37--17.42~\AA\ range,
a few others at 17.447~\AA, and several others in the range
17.494--17.510~\AA, i.e. effectively the laboratory spectrum
has the same lines as the solar one (including a blend of
emission for the first two lines). 
Our model also predicts many lines mostly formed by
DC, but none that stand out  for their brightness, so identifications are
very difficult.

The brightest is a decay from  3s 3p $^4$D$_{7/2}$ (level No.38)
to the 3p $^2$P$_{3/2}$, predicted at 17.63~\AA.
G09  has a similar predicted wavelength
(17.619~\AA), but gives an observed wavelength of 17.592~\AA,
which does not agree with the solar spectrum. D13 does not provide a calculated
energy for the upper state, nor B76.

Similarly, there are several lines from low-lying states that our model predicts
around 17.55~\AA. The strongest of them, the decay of the 3s 3d $^4$P$_{5/2}$
to the 3d $^2$D$_{5/2}$, has a wavelength predicted by  \cite{liang_etal:2008} of 
17.36~\AA, but 
was instead identified by B76 with a line at 17.498~\AA.
Our identification agrees with G09.

\section{Conclusions}

Despite being  a century from their discovery, satellite lines are still
relatively poorly known.
The literature on the identifications of the strongest lines is very confusing.
Proper modelling requires large-scale calculations
of accurate atomic data and  further studies.
From our brief summary of previous studies, both theoretical and experimental, it is clear
that, especially  for the satellite lines of Na-like iron, further 
studies are needed to benchmark the atomic data and to obtain experimental wavelengths.
It is also clear that different  theoretical approaches can provide significantly different
models.

Such  work is particularly important for Fe XVI as the satellite lines can be relatively
strong and blend the   Fe XVII 3C and 3D lines, which are among the most important X-ray lines,
because of  their plasma diagnostic use.
We found that the missing flux around the 3C and 3D lines, about a factor of 2 as found
from the analysis of the first MaGIXS flight, is mostly due to decays from autoionizing states
of Fe XVI. 

We have also identified for the first time
many  other lines in the X-rays, and showed that some are also blending
previously known lines.
It is clear that a complete knowledge of these satellite lines is important when analysing
astrophysical observations at these wavelengths, for example from the
recently launched XRISM satellite. It is also clear that further calculations
for other ions are needed.

Our comparison to the best available solar high-resolution spectra is very
satisfactory, but further observations are needed to test the models.
In this respect, the current NASA proposals for further sounding rockets with an
improved  MaGIXS instrument  are very important.

\section*{Acknowledgments}
GDZ acknowledges support from STFC (UK) via the consolidated grants 
to the atomic astrophysics group (AAG) at DAMTP, University of Cambridge (ST/P000665/1. and ST/T000481/1).
The UK APAP network  is funded by STFC via the consolidated grant (PI Badnell) to the
University of Strathclyde (ST/V000683/1).

\section*{Data Availability}
The results of the largest, fine-structure resolved calculation are made available
at ZENODO in CHIANTI ascii format, for easy inclusion.

\bibliographystyle{mn2e}

\bibliography{paper.bib}

\clearpage

\appendix

\section{Additional material}

\begin{table}
\caption{The target electron configuration basis 
 and orbital scaling parameters $\lambda_{nl}$ 
for the structure  run of Liang et al.
}
\begin{flushleft}
\begin{tabular}{l|llllllllllllllll}
\hline\hline\noalign{\smallskip}
Configurations     &               &  &       \\ 
\noalign{\smallskip}\hline\noalign{\smallskip}
 1s$^2$ 2s$^2$ 2p$^6$ 3$l$  ($l$=s,p,d)       &   1s & 1.39364 &      \\
 1s$^2$ 2s$^2$ 2p$^5$ 3$l$ 3$l'$  ($l$=s,p,d, $l'$=s,p ) &   2s & 1.08686  & 3s  & 1.15588 \\
                                              &  2p & 1.02341 & 3p  &  1.11371  \\ 
                                              & 3d &  1.15100 &    & \\ 
 \noalign{\smallskip}\hline
\end{tabular}
\normalsize
\end{flushleft}
\label{tab:basis}
\end{table}

\begin{figure}
\centerline{\epsfig{file=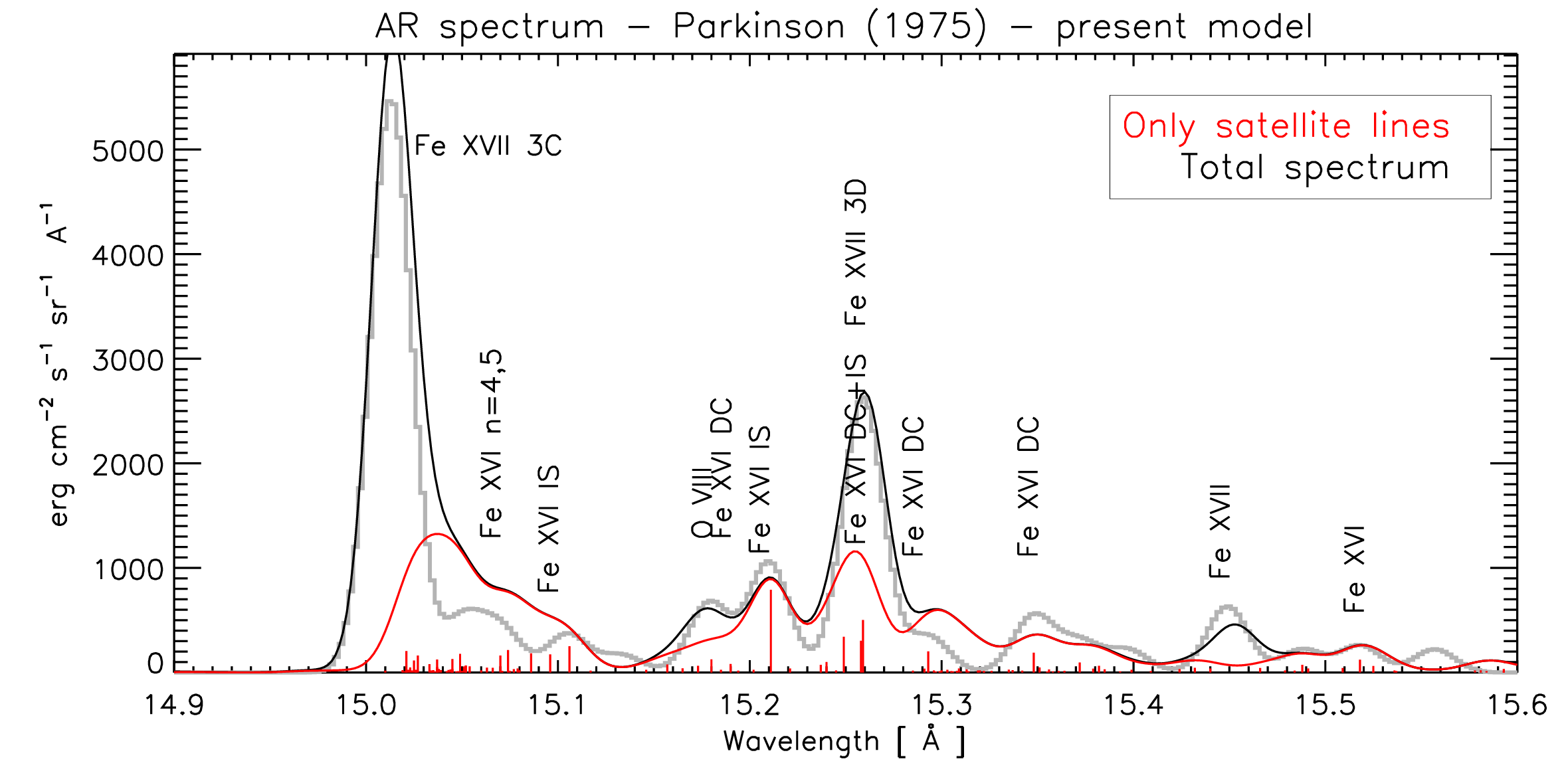, width=9.cm,angle=0 }}
\centerline{\epsfig{file=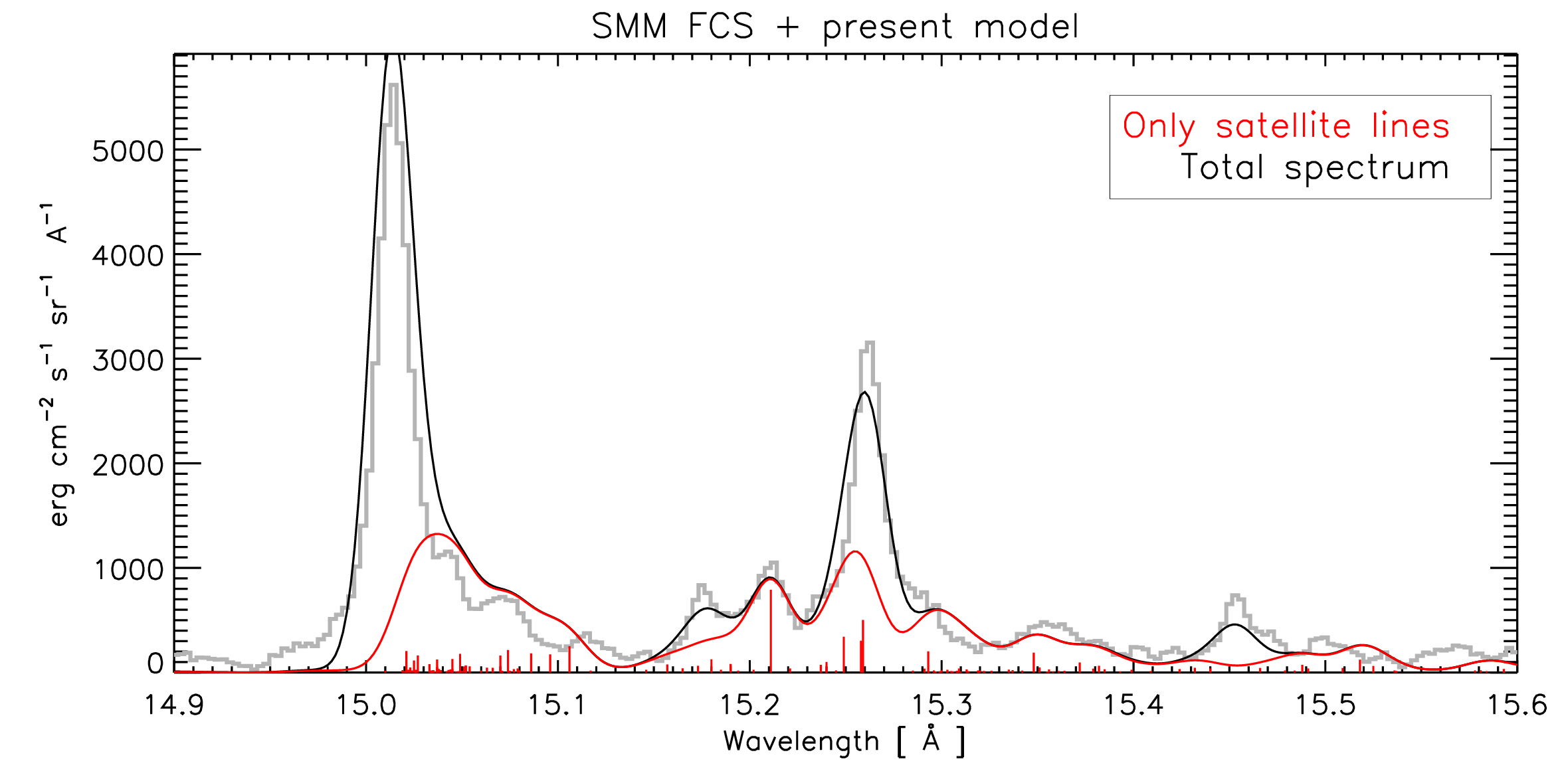, width=9.cm,angle=0 }}
\caption{A comparison between the reconstructed P75 spectrum (grey thick line) and an
averaged SMM/FCS spectrum of an active region with the 
 present $n=6$ model
in the 15-15.6~\AA\ region.  }
\label{fig:sp15_n6}
\end{figure}

\begin{table}
\begin{center}
  \caption{List of the main states.
    \label{tab:table_e_n10} }
\footnotesize
\setlength\tabcolsep{3.5pt} 
\begin{tabular}{@{}rllllllllllllllllllllll@{}}
  \hline\noalign{\smallskip}
 $i$ &  Conf.  & P & T & $E_{\rm exp}$ &  $E_{\rm AS}$  & $E_{\rm Diaz+}$ & $E_{\rm Liang+}$  \\ 
\noalign{\smallskip}
  \hline
    1 &         2s$^{2}$ 2p$^{6}$ 3s &   e &   $^2$S$_{1/2}$ &         0 &         0 &         0 &         0 & \\ 
    2 &         2s$^{2}$ 2p$^{6}$ 3p &   o &   $^2$P$_{1/2}$ &    277194 &    276293 &    277222 &    276436 & \\ 
    3 &         2s$^{2}$ 2p$^{6}$ 3p &   o &   $^2$P$_{3/2}$ &    298143 &    298373 &    298167 &    296534 & \\ 
    4 &         2s$^{2}$ 2p$^{6}$ 3d &   e &   $^2$D$_{3/2}$ &    675501 &    674642 &    675463 &    676373 & \\ 
    5 &         2s$^{2}$ 2p$^{6}$ 3d &   e &   $^2$D$_{5/2}$ &    678405 &    679275 &    678372 &    679712 & \\ 

  \noalign{\smallskip}
   77 &   2s$^{2}$ 2p$^{5}$ 3s$^{2}$ &   o &   $^2$P$_{3/2}$ &   5773000? &   5748000 &   5756556 &   5802584 & \\ 
   78 &   2s$^{2}$ 2p$^{5}$ 3s$^{2}$ &   o &   $^2$P$_{1/2}$ &   5873000? &   5850943 &   5857665 &   5899697 & \\ 
   79 &      2s$^{2}$ 2p$^{5}$ 3s 3p &   e &   $^4$S$_{3/2}$ &  -   &   5941810 &   5953391 &   5991935 & \\ 
   80 &      2s$^{2}$ 2p$^{5}$ 3s 3p &   e &   $^4$D$_{5/2}$ &   5982000 &   5970059 &   5980479 &   6020272 & \\ 
   81 &      2s$^{2}$ 2p$^{5}$ 3s 3p &   e &   $^4$D$_{7/2}$ &  -   &   5976391 &   5986775 &   6026148 & \\ 
   82 &      2s$^{2}$ 2p$^{5}$ 3s 3p &   e &   $^2$P$_{3/2}$ &  -   &   5977178 &   5987047 &   6027021 & \\ 
   83 &      2s$^{2}$ 2p$^{5}$ 3s 3p &   e &   $^2$P$_{1/2}$ &   6001000 &   5989400 &   5999543 &   6041011 & \\ 
   84 &      2s$^{2}$ 2p$^{5}$ 3s 3p &   e &   $^4$P$_{5/2}$ &   6013000 &   6001435 &   6011855 &   6053544 & \\ 
   85 &      2s$^{2}$ 2p$^{5}$ 3s 3p &   e &   $^2$D$_{3/2}$ &   6013000 &   6002818 &   6012375 &   6053898 & \\ 
   86 &      2s$^{2}$ 2p$^{5}$ 3s 3p &   e &   $^2$S$_{1/2}$ &   6042000? &   6019309 &   6027754 &   6076536 & \\ 
   87 &      2s$^{2}$ 2p$^{5}$ 3s 3p &   e &   $^4$D$_{1/2}$ &   6075000 &   6068977 &   6077192 &   6113566 & \\ 
   88 &      2s$^{2}$ 2p$^{5}$ 3s 3p &   e &   $^4$P$_{1/2}$ &   6089000? &   6074866 &   6082835 &   6128206 & \\ 
   89 &      2s$^{2}$ 2p$^{5}$ 3s 3p &   e &   $^4$D$_{3/2}$ &   6089000 &   6079784 &   6087509 &   6124285 & \\ 
   90 &      2s$^{2}$ 2p$^{5}$ 3s 3p &   e &   $^2$D$_{5/2}$ &  -   &   6090463 &   6096282 &   6141431 & \\ 
   91 &      2s$^{2}$ 2p$^{5}$ 3s 3p &   e &   $^4$P$_{3/2}$ &   6096000 &   6091914 &   6100268 &   6138528 & \\ 
   92 &      2s$^{2}$ 2p$^{5}$ 3s 3p &   e &   $^2$D$_{5/2}$ &   6110000 &   6101305 &   6108077 &   6147237 & \\ 
   93 &      2s$^{2}$ 2p$^{5}$ 3s 3p &   e &   $^2$P$_{3/2}$ &   6129000 &   6108503 &   6113831 &   6157761 & \\ 
   94 &      2s$^{2}$ 2p$^{5}$ 3s 3p &   e &   $^2$P$_{1/2}$ &      -   &   6181158 &   6182346 &   6229457 & \\ 
   95 &      2s$^{2}$ 2p$^{5}$ 3s 3p &   e &   $^2$D$_{3/2}$ &   6217000? &   6198457 &   6201702 &   6244142 & \\ 
   96 &      2s$^{2}$ 2p$^{5}$ 3s 3p &   e &   $^2$S$_{1/2}$ &   6267000? &   6253611 &   6245187 &   6313279 & \\ 
   
\noalign{\smallskip}
  111 &      2s$^{2}$ 2p$^{5}$ 3s 3d &   o &   $^4$P$_{5/2}$ &   6393000 &   6382976 &   6390567 &   6440048 & \\ 
  112 &      2s$^{2}$ 2p$^{5}$ 3s 3d &   o &   $^4$F$_{9/2}$ &  -   &   6383060 &   6389221 &   6438162 & \\ 
\noalign{\smallskip}
  117 &      2s$^{2}$ 2p$^{5}$ 3s 3d &   o &   $^4$F$_{5/2}$ &   6406000 &   6397726 &   6404701 &   6453145 & \\ 
  118 &   2s$^{2}$ 2p$^{5}$ 3p$^{2}$ &   o &   $^2$P$_{3/2}$ &  -   &   6400493 &   6406003 &   6469670 & \\ 
  119 &      2s$^{2}$ 2p$^{5}$ 3s 3d &   o &   $^2$D$_{3/2}$ &   6419000 &   6408540 &   6415660 &   6464402 & \\ 
  120 &   2s$^{2}$ 2p$^{5}$ 3p$^{2}$ &   o &   $^2$D$_{3/2}$ &  -   &   6415643 &   6422064 &   6455698 & \\ 
  121 &      2s$^{2}$ 2p$^{5}$ 3s 3d &   o &   $^4$D$_{7/2}$ &   6422000 &   6416259 &   6421329 &   6471649 & \\ 
  122 &   2s$^{2}$ 2p$^{5}$ 3p$^{2}$ &   o &   $^2$D$_{5/2}$ &   -        &   6418038 &   6423498 &   6464730 & \\ 
  123 &      2s$^{2}$ 2p$^{5}$ 3s 3d &   o &   $^2$P$_{1/2}$ &  6423000   &   6418066 &   6423578 &   6476262 & \\ 
  124 &      2s$^{2}$ 2p$^{5}$ 3s 3d &   o &   $^2$F$_{5/2}$ &   6425000 &   6420114 &   6425339 &   6474699 & \\
  125 &      2s$^{2}$ 2p$^{5}$ 3s 3d &   o &   $^2$P$_{3/2}$ &   \sout{6436000} &   6439623 &   6443091 &   6498398 & \\
     & & & & {\bf 6444100?} & & & \\ 
  126 &      2s$^{2}$ 2p$^{5}$ 3s 3d &   o &   $^4$D$_{1/2}$ &  -   &   6450153 &   6455202 &   6506414 & \\
  127 &      2s$^{2}$ 2p$^{5}$ 3s 3d &   o &   $^4$D$_{3/2}$ &   \sout{6473000}   &   6479009 &   6483365 &   6536053 & \\ 
  128 &      2s$^{2}$ 2p$^{5}$ 3s 3d &   o &   $^2$F$_{7/2}$ &  \sout{6445000}  &   6483824 &   6485011 &   6546990 & \\ 
  129 &      2s$^{2}$ 2p$^{5}$ 3s 3d &   o &   $^4$F$_{3/2}$ &   6502000 &   6496562 &   6502061 &   6547383 & \\ 
  130 &      2s$^{2}$ 2p$^{5}$ 3s 3d &   o &   $^2$D$_{5/2}$ &  \sout{6464000}     &   6498874 &   6501608 &   6555370 & \\ 
  131 &      2s$^{2}$ 2p$^{5}$ 3s 3d &   o &   $^4$D$_{5/2}$ &   6502000 &   6499710 &   6504077 &   6549706 & \\ 

\noalign{\smallskip}
\hline
\end{tabular}
\begin{tablenotes}
    \item[] { $E_{\rm exp}$ gives the NIST experimental energies, except the
   those in bold which are our tentative values. $E_{\rm AS}$ are our ab-initio
   AS energies with the $n=10$ model. $E_{\rm Diaz+}$ are the energies from \cite{diaz_etal:2013}
while  $E_{\rm Liang+}$ are the AS ones from \cite{liang_etal:2008}. }
 \end{tablenotes}
\end{center}
\normalsize
\end{table}

\addtocounter{table}{-1}
\begin{table}
\begin{center}
   \caption{Contd  }
\footnotesize
\setlength\tabcolsep{3.5pt} 
\begin{tabular}{@{}rllllllllllllllllllllll@{}}
  \hline\noalign{\smallskip}
 $i$ &  Conf.  & P & T & $E_{\rm NIST}$ &  $E_{\rm AS}$  & $E_{\rm Diaz+}$ & $E_{\rm Liang+}$  \\ 
\noalign{\smallskip}
  \hline
 \noalign{\smallskip}
  132 &   2s$^{2}$ 2p$^{5}$ 3p$^{2}$ &   o &   $^2$P$_{1/2}$ &  -   &   6507356 &   6508883 &   6566725 & \\ 
  133 &      2s$^{2}$ 2p$^{5}$ 3s 3d &   o &   $^2$D$_{5/2}$ &   6516000 &   6511948 &   6514575 &   6569938 & \\ 
  134 &      2s$^{2}$ 2p$^{5}$ 3s 3d &   o &   $^2$F$_{7/2}$ &   6517000 &   6512092 &   6514871 &   6561652 & \\ 
  135 &   2s$^{2}$ 2p$^{5}$ 3p$^{2}$ &   o &   $^2$P$_{1/2}$ &  -   &   6512966 &   6514341 &   6579688 & \\ 
  136 &   2s$^{2}$ 2p$^{5}$ 3p$^{2}$ &   o &   $^2$P$_{3/2}$ &  -   &   6530425 &   6531608 &   6592253 & \\ 
  137 &      2s$^{2}$ 2p$^{5}$ 3s 3d &   o &   $^2$D$_{3/2}$ &   \sout{6530000}   &   6551020 &   6550184 &   6611638 & \\
      &                             &     &                &  {\bf 6553500?}   &        &        &      \\ 
  138 &      2s$^{2}$ 2p$^{5}$ 3s 3d &   o &   $^2$P$_{1/2}$ &   6574000 &   6576810 &   6573657 &   6644694 & \\ 
  139 &      2s$^{2}$ 2p$^{5}$ 3p 3d &   e &   $^4$D$_{1/2}$ &  -   &   6589929 &   6601400 &   6646599 & \\ 
  140 &      2s$^{2}$ 2p$^{5}$ 3s 3d &   o &   $^2$F$_{5/2}$ &  \sout{6556000}   &   6594116 &   6593543 &   6651977 & \\ 
  141 &      2s$^{2}$ 2p$^{5}$ 3p 3d &   e &   $^4$D$_{3/2}$ &  -   &   6598110 &   6608991 &   6654887 & \\ 
  142 &      2s$^{2}$ 2p$^{5}$ 3p 3d &   e &   $^4$D$_{5/2}$ &  -   &   6610901 &   6620899 &   6667902 & \\ 
  143 &      2s$^{2}$ 2p$^{5}$ 3s 3d &   o &   $^2$P$_{3/2}$ &  \sout{6595000}  &   6619300 &   6616740 &   6686516 & \\ 
      &                             &     &                &   {\bf 6620000?} &        &        &      \\
\noalign{\smallskip}
  195 &      2s$^{2}$ 2p$^{5}$ 3p 3d &   e &   $^2$D$_{3/2}$ &  {\bf 6831000?} &   6834145 &   6831282 &   6894915 & \\ 
  196 &      2s$^{2}$ 2p$^{5}$ 3p 3d &   e &   $^2$D$_{5/2}$ &  {\bf 6837100?} &   6839169 &   6838045 &   6893483 & \\ 
\noalign{\smallskip}
  235 &   2s$^{2}$ 2p$^{5}$ 3d$^{2}$ &   o &   $^2$G$_{7/2}$ &  {\bf 7135000?} &   7132714 &   7134361 &  -   & \\ 
  236 &   2s$^{2}$ 2p$^{5}$ 3d$^{2}$ &   o &   $^4$F$_{5/2}$ &  -   &   7139463 &   7142053 &  -   & \\ 
  237 &            2s 2p$^{6}$ 3s 3p &   o &   $^2$P$_{3/2}$ &  -   &   7140998 &   7128949 &  -   & \\ 
\noalign{\smallskip}
  244 &   2s$^{2}$ 2p$^{5}$ 3d$^{2}$ &   o &   $^2$F$_{5/2}$ &  {\bf 7180800?} &   7185725 &   7186088 &  -   & \\ 
  245 &   2s$^{2}$ 2p$^{5}$ 3d$^{2}$ &   o &   $^2$F$_{5/2}$ &  {\bf 7191000?} &   7190776 &   7193832 &  -   & \\ 

\noalign{\smallskip}
  254 &   2s$^{2}$ 2p$^{5}$ 3d$^{2}$ &   o &   $^2$F$_{7/2}$ &  {\bf  7236000?} &   7243251 &   7242818 &  -   & \\ 
  255 &   2s$^{2}$ 2p$^{5}$ 3d$^{2}$ &   o &   $^2$D$_{5/2}$ &  {\bf  7240000?} &   7252103 &   7246119 &  -   & \\ 
  256 &   2s$^{2}$ 2p$^{5}$ 3d$^{2}$ &   o &   $^2$D$_{3/2}$ &  -   &   7258926 &   7253388 &  -   & \\ 
  257 &   2s$^{2}$ 2p$^{5}$ 3d$^{2}$ &   o &   $^2$P$_{3/2}$ &  {\bf  7266000?} &   7270449 &   7263947 &  -   & \\ 

\noalign{\smallskip}
\hline
\end{tabular}
\end{center}
\normalsize
\end{table}

\clearpage
\onecolumn

\input{comp_el_publ}

\clearpage
\twocolumn

\end{document}

%% file: comp_el_publ.tex
\footnotesize

\begin{longtable}[c]{@{}|rlclcllll|@{}}
  \caption{ Energies \label{tab:energies}} \\
\hline
 $i$ &  Conf.  & P &  T & $E_{\rm NIST}$ &  $E_{AS}$ & $E_{\rm Diaz+}$ & $E_{\rm Liang+}$ &  \\ \hline
\hline \endfirsthead
\caption[]{\normalsize (continued)}\\ \hline
 $i$ &  Conf.  & P &  T & $E_{\rm NIST}$ &  $E_{AS}$ & $E_{\rm Diaz+}$ & $E_{\rm Liang+}$ &  \\ \hline
\hline \endhead
    1 &         2s$^{2}$ 2p$^{6}$ 3s &   e &   $^2$S$_{1/2}$ &         0 &         0 &         0 &         0 & \\ 
    2 &         2s$^{2}$ 2p$^{6}$ 3p &   o &   $^2$P$_{1/2}$ &    277194 &    277711 &    277222 &    276436 & \\ 
    3 &         2s$^{2}$ 2p$^{6}$ 3p &   o &   $^2$P$_{3/2}$ &    298143 &    300089 &    298167 &    296534 & \\ 
    4 &         2s$^{2}$ 2p$^{6}$ 3d &   e &   $^2$D$_{3/2}$ &    675501 &    676579 &    675463 &    676373 & \\ 
    5 &         2s$^{2}$ 2p$^{6}$ 3d &   e &   $^2$D$_{5/2}$ &    678405 &    681330 &    678372 &    679712 & \\ 
    6 &         2s$^{2}$ 2p$^{6}$ 4s &   e &   $^2$S$_{1/2}$ &   1867740 &   1869087 &   1867664 &   1867895 & \\ 
    7 &         2s$^{2}$ 2p$^{6}$ 4p &   o &   $^2$P$_{1/2}$ &   1977649 &   1979132 &   1977616 &   1977070 & \\ 
    8 &         2s$^{2}$ 2p$^{6}$ 4p &   o &   $^2$P$_{3/2}$ &   1985649 &   1987870 &   1985786 &   1984703 & \\ 
    9 &         2s$^{2}$ 2p$^{6}$ 4d &   e &   $^2$D$_{3/2}$ &   2124719 &   2126589 &   2124584 &   2124092 & \\ 
   10 &         2s$^{2}$ 2p$^{6}$ 4d &   e &   $^2$D$_{5/2}$ &   2125959 &   2128787 &   2125923 &   2125524 & \\ 
   11 &         2s$^{2}$ 2p$^{6}$ 4f &   o &   $^2$F$_{5/2}$ &   2184960 &   2184613 &   2184910 &   2184919 & \\ 
   12 &         2s$^{2}$ 2p$^{6}$ 4f &   o &   $^2$F$_{7/2}$ &   2185409 &   2185387 &   2185401 &   2185424 & \\ 
   13 &         2s$^{2}$ 2p$^{6}$ 5s &   e &   $^2$S$_{1/2}$ &   2662000 &   2663585 &   2663328 &   2662751 & \\ 
   14 &         2s$^{2}$ 2p$^{6}$ 5p &   o &   $^2$P$_{1/2}$ &   2717169 &   2717783 &   2717620 &   2716785 & \\ 
   15 &         2s$^{2}$ 2p$^{6}$ 5p &   o &   $^2$P$_{3/2}$ &   2721159 &   2722230 &   2721636 &   2720472 & \\ 
   16 &         2s$^{2}$ 2p$^{6}$ 5d &   e &   $^2$D$_{3/2}$ &   2788049 &   2789109 &   2788713 &   2787715 & \\ 
   17 &         2s$^{2}$ 2p$^{6}$ 5d &   e &   $^2$D$_{5/2}$ &   2788609 &   2790271 &   2789416 &   2788448 & \\ 
   18 &         2s$^{2}$ 2p$^{6}$ 5f &   o &   $^2$F$_{5/2}$ &   2818599 &   2818306 &   2818974 &   2818342 & \\ 
   19 &         2s$^{2}$ 2p$^{6}$ 5f &   o &   $^2$F$_{7/2}$ &   2818900 &   2818714 &   2819226 &   2818601 & \\ 
   20 &         2s$^{2}$ 2p$^{6}$ 5g &   e &   $^2$G$_{7/2}$ &   2822700 &   2821171 &         0 &   2821317 & \\ 
   21 &         2s$^{2}$ 2p$^{6}$ 5g &   e &   $^2$G$_{9/2}$ &   2822800 &   2821402 &         0 &   2821470 & \\ 
   22 &         2s$^{2}$ 2p$^{6}$ 6s &   e &   $^2$S$_{1/2}$ &   3075999 &   3075690 &         0 &   3075119 & \\ 
   23 &         2s$^{2}$ 2p$^{6}$ 6p &   o &   $^2$P$_{1/2}$ &   3106400 &   3106229 &         0 &   3105599 & \\ 
   24 &         2s$^{2}$ 2p$^{6}$ 6p &   o &   $^2$P$_{3/2}$ &   3108899 &   3108902 &  -   &   3107635 & \\ 
   25 &         2s$^{2}$ 2p$^{6}$ 6d &   e &   $^2$D$_{3/2}$ &   3146070 &   3146303 &  -   &   3145288 & \\ 
   26 &         2s$^{2}$ 2p$^{6}$ 6d &   e &   $^2$D$_{5/2}$ &   3146670 &   3147021 &  -   &   3145709 & \\ 
   27 &         2s$^{2}$ 2p$^{6}$ 6f &   o &   $^2$F$_{5/2}$ &   3163129 &   3162874 &  -   &   3162816 & \\ 
   28 &         2s$^{2}$ 2p$^{6}$ 6f &   o &   $^2$F$_{7/2}$ &   3163190 &   3163122 &  -   &   3162966 & \\ 
   29 &         2s$^{2}$ 2p$^{6}$ 6g &   e &   $^2$G$_{7/2}$ &  -   &   3164672 &  -   &   3164757 & \\ 
   30 &         2s$^{2}$ 2p$^{6}$ 6g &   e &   $^2$G$_{9/2}$ &  -   &   3164807 &  -   &   3164846 & \\ 
   31 &         2s$^{2}$ 2p$^{6}$ 6h &   o &   $^2$H$_{9/2}$ &  -   &   3164858 &  -   &   3164916 & \\ 
   32 &         2s$^{2}$ 2p$^{6}$ 6h &   o &   $^2$H$_{11/2}$ &  -   &   3164947 &  -   &   3164976 & \\ 

  33 &   2s$^{2}$ 2p$^{5}$ 3s$^{2}$ &   o &   $^2$P$_{3/2}$ &   5773000? &   5744641 &   5756556 &   5802584 & \\ 
   34 &   2s$^{2}$ 2p$^{5}$ 3s$^{2}$ &   o &   $^2$P$_{1/2}$ &   5873000? &   5848114 &   5857665 &   5899697 & \\ 
   35 &      2s$^{2}$ 2p$^{5}$ 3s 3p &   e &   $^4$S$_{3/2}$ &  -   &   5939043 &   5953391 &   5991935 & \\ 
   36 &      2s$^{2}$ 2p$^{5}$ 3s 3p &   e &   $^4$D$_{5/2}$ &   5982000 &   5967095 &   5980479 &   6020272 & \\ 
   37 &      2s$^{2}$ 2p$^{5}$ 3s 3p &   e &   $^4$D$_{7/2}$ &  -   &   5973428 &   5986775 &   6026148 & \\ 
   38 &      2s$^{2}$ 2p$^{5}$ 3s 3p &   e &   $^2$P$_{3/2}$ &  -   &   5974184 &   5987047 &   6027021 & \\ 
   39 &      2s$^{2}$ 2p$^{5}$ 3s 3p &   e &   $^2$P$_{1/2}$ &   6001000 &   5986456 &   5999543 &   6041011 & \\ 
   40 &      2s$^{2}$ 2p$^{5}$ 3s 3p &   e &   $^4$P$_{5/2}$ &   6013000 &   5998400 &   6011855 &   6053544 & \\ 
   41 &      2s$^{2}$ 2p$^{5}$ 3s 3p &   e &   $^2$D$_{3/2}$ &   6013000 &   5999767 &   6012375 &   6053898 & \\ 
   42 &      2s$^{2}$ 2p$^{5}$ 3s 3p &   e &   $^2$S$_{1/2}$ &   6042000? &   6016544 &   6027754 &   6076536 & \\ 
   43 &      2s$^{2}$ 2p$^{5}$ 3s 3p &   e &   $^4$D$_{1/2}$ &   6075000 &   6066510 &   6077192 &   6113566 & \\ 
   44 &      2s$^{2}$ 2p$^{5}$ 3s 3p &   e &   $^4$P$_{1/2}$ &   6089000? &   6072233 &   6082835 &   6128206 & \\ 
   45 &      2s$^{2}$ 2p$^{5}$ 3s 3p &   e &   $^4$D$_{3/2}$ &   6089000 &   6077288 &   6087509 &   6124285 & \\ 
   46 &      2s$^{2}$ 2p$^{5}$ 3s 3p &   e &   $^2$D$_{5/2}$ &       -   &   6087412 &   6096282 &   6141431 & \\ 
   47 &      2s$^{2}$ 2p$^{5}$ 3s 3p &   e &   $^4$P$_{3/2}$ &   6096000 &   6089389 &   6100268 &   6138528 & \\ 
   48 &      2s$^{2}$ 2p$^{5}$ 3s 3p &   e &   $^2$D$_{5/2}$ &   6110000 &   6098693 &   6108077 &   6147237 & \\ 
   49 &      2s$^{2}$ 2p$^{5}$ 3s 3p &   e &   $^2$P$_{3/2}$ &   6129000? &   6105380 &   6113831 &   6157761 & \\ 
   50 &      2s$^{2}$ 2p$^{5}$ 3s 3p &   e &   $^2$P$_{1/2}$ &  -   &   6178837 &   6182346 &   6229457 & \\ 
   51 &      2s$^{2}$ 2p$^{5}$ 3s 3p &   e &   $^2$D$_{3/2}$ &   6217000? &   6195854 &   6201702 &   6244142 & \\ 
   52 &      2s$^{2}$ 2p$^{5}$ 3s 3p &   e &   $^2$S$_{1/2}$ &   6267000? &   6252510 &   6245187 &   6313279 & \\ 

   53 &   2s$^{2}$ 2p$^{5}$ 3p$^{2}$ &   o &   $^4$P$_{3/2}$ &  -   &   6257463 &   6269659 &   6316508 & \\ 
   54 &   2s$^{2}$ 2p$^{5}$ 3p$^{2}$ &   o &   $^2$P$_{1/2}$ &  -   &   6258792 &   6271517 &   6317822 & \\ 
   55 &   2s$^{2}$ 2p$^{5}$ 3p$^{2}$ &   o &   $^4$P$_{5/2}$ &  -   &   6265662 &   6278181 &   6324087 & \\ 
   56 &   2s$^{2}$ 2p$^{5}$ 3p$^{2}$ &   o &   $^2$F$_{7/2}$ &  -   &   6276200 &   6287623 &   6328785 & \\ 
   57 &   2s$^{2}$ 2p$^{5}$ 3p$^{2}$ &   o &   $^2$P$_{3/2}$ &  -   &   6278747 &   6291100 &   6334949 & \\ 
   58 &   2s$^{2}$ 2p$^{5}$ 3p$^{2}$ &   o &   $^4$P$_{1/2}$ &  -   &   6295965 &   6307896 &   6353532 & \\ 
   59 &   2s$^{2}$ 2p$^{5}$ 3p$^{2}$ &   o &   $^2$D$_{5/2}$ &  -   &   6297314 &   6308932 &   6350782 & \\ 
   60 &   2s$^{2}$ 2p$^{5}$ 3p$^{2}$ &   o &   $^2$D$_{3/2}$ &  -   &   6298455 &   6308901 &   6352715 & \\ 
   61 &   2s$^{2}$ 2p$^{5}$ 3p$^{2}$ &   o &   $^4$D$_{7/2}$ &  -   &   6303014 &   6314133 &   6356554 & \\ 
   62 &   2s$^{2}$ 2p$^{5}$ 3p$^{2}$ &   o &   $^4$D$_{5/2}$ &  -   &   6304417 &   6315498 &   6358568 & \\ 
   63 &   2s$^{2}$ 2p$^{5}$ 3p$^{2}$ &   o &   $^4$D$_{1/2}$ &  -   &   6353354 &   6362006 &   6404267 & \\ 
   64 &      2s$^{2}$ 2p$^{5}$ 3s 3d &   o &   $^4$P$_{1/2}$ &  -   &   6356616 &   6369713 &   6418318 & \\ 
   65 &   2s$^{2}$ 2p$^{5}$ 3p$^{2}$ &   o &   $^4$S$_{3/2}$ &  -   &   6357933 &   6368079 &   6411844 & \\ 
   66 &      2s$^{2}$ 2p$^{5}$ 3s 3d &   o &   $^4$P$_{3/2}$ &  -   &   6365097 &   6377262 &   6426196 & \\ 
   67 &      2s$^{2}$ 2p$^{5}$ 3s 3d &   o &   $^4$P$_{5/2}$ &   6393000 &   6379612 &   6390567 &   6440048 & \\ 
   68 &      2s$^{2}$ 2p$^{5}$ 3s 3d &   o &   $^4$F$_{9/2}$ &  -   &   6379790 &   6389221 &   6438162 & \\ 
   69 &   2s$^{2}$ 2p$^{5}$ 3p$^{2}$ &   o &   $^4$D$_{3/2}$ &  -   &   6381562 &   6390640 &   6432092 & \\ 
   70 &   2s$^{2}$ 2p$^{5}$ 3p$^{2}$ &   o &   $^2$F$_{5/2}$ &  -   &   6385341 &   6394058 &   6431790 & \\ 
   71 &      2s$^{2}$ 2p$^{5}$ 3s 3d &   o &   $^4$F$_{7/2}$ &  -   &   6385866 &   6396084 &   6444453 & \\ 
   72 &   2s$^{2}$ 2p$^{5}$ 3p$^{2}$ &   o &   $^2$S$_{1/2}$ &  -   &   6390585 &   6398771 &   6439224 & \\ 
   73 &      2s$^{2}$ 2p$^{5}$ 3s 3d &   o &   $^4$F$_{5/2}$ &   6406000 &   6394429 &   6404701 &   6453145 & \\ 
   74 &   2s$^{2}$ 2p$^{5}$ 3p$^{2}$ &   o &   $^2$P$_{3/2}$ &  -   &   6397961 &   6406003 &   6469670 & \\ 
   75 &      2s$^{2}$ 2p$^{5}$ 3s 3d &   o &   $^2$D$_{3/2}$ &   6419000 &   6405385 &   6415660 &   6464402 & \\ 
   76 &   2s$^{2}$ 2p$^{5}$ 3p$^{2}$ &   o &   $^2$D$_{3/2}$ &  -   &   6413008 &   6422064 &   6455698 & \\ 
   77 &      2s$^{2}$ 2p$^{5}$ 3s 3d &   o &   $^4$D$_{7/2}$ &   6422000 &   6413123 &   6421329 &   6471649 & \\ 
   78 &      2s$^{2}$ 2p$^{5}$ 3s 3d &   o &   $^2$P$_{1/2}$ &  -   &   6415027 &   6423498 &   6464730 & \\ 
   79 &      2s$^{2}$ 2p$^{5}$ 3s 3d &   o &   $^2$F$_{5/2}$ &   6423000 &   6415181 &   6423578 &   6476262 & \\ 
   80 &   2s$^{2}$ 2p$^{5}$ 3p$^{2}$ &   o &   $^2$D$_{5/2}$ &   6425000 &   6417371 &   6425339 &   6474699 & \\ 

  81 &      2s$^{2}$ 2p$^{5}$ 3s 3d &   o &   $^2$P$_{3/2}$ &  \sout{6436000}  &   6436676 &   6443091 &   6498398 & \\
      & & & & {\bf 6444100?} & & & \\
   82 &      2s$^{2}$ 2p$^{5}$ 3s 3d &   o &   $^4$D$_{1/2}$ &              -   &   6447867 &   6455202 &   6506414 & \\ 
   83 &      2s$^{2}$ 2p$^{5}$ 3s 3d &   o &   $^4$D$_{3/2}$ &  \sout{6473000}   &   6476481 &   6483365 &   6536053 & \\ 
   84 &      2s$^{2}$ 2p$^{5}$ 3s 3d &   o &   $^2$F$_{7/2}$ &  \sout{6445000}   &   6480673 &   6485011 &   6546990 & \\ 
   85 &      2s$^{2}$ 2p$^{5}$ 3s 3d &   o &   $^4$F$_{3/2}$ &   6502000 &   6493786 &   6502061 &   6547383 & \\ 
   86 &      2s$^{2}$ 2p$^{5}$ 3s 3d &   o &   $^2$D$_{5/2}$ &  \sout{6464000}   &   6495941 &   6501608 &   6549706 & \\ 
   87 &      2s$^{2}$ 2p$^{5}$ 3s 3d &   o &   $^4$D$_{5/2}$ &   6502000 &   6496993 &   6504077 &   6555370 & \\
   88 &   2s$^{2}$ 2p$^{5}$ 3p$^{2}$ &   o &   $^2$P$_{1/2}$ &  -   &   6504938 &   6508883 &   6566725 & \\ 
   89 &      2s$^{2}$ 2p$^{5}$ 3s 3d &   o &   $^2$D$_{5/2}$ &   6516000 &   6508973 &   6514575 &   6569938 & \\ 
   90 &      2s$^{2}$ 2p$^{5}$ 3s 3d &   o &   $^2$F$_{7/2}$ &   6517000 &   6509407 &   6514871 &   6561652 & \\ 
   91 &   2s$^{2}$ 2p$^{5}$ 3p$^{2}$ &   o &   $^2$P$_{1/2}$ &  -   &   6511585 &   6514341 &   6579688 & \\ 
   92 &   2s$^{2}$ 2p$^{5}$ 3p$^{2}$ &   o &   $^2$P$_{3/2}$ &  -   &   6528784 &   6531608 &   6592253 & \\ 
   93 &      2s$^{2}$ 2p$^{5}$ 3s 3d &   o &   $^2$D$_{3/2}$ &   \sout{6530000} &   6549199 &   6550184 &   6611638 & \\
        &                             &     &                &  {\bf 6553500?}   &        &        &      \\ 

   94 &      2s$^{2}$ 2p$^{5}$ 3s 3d &   o &   $^2$P$_{1/2}$ &   6574000 &   6575409 &   6573657 &   6644694 & \\ 
   95 &      2s$^{2}$ 2p$^{5}$ 3p 3d &   e &   $^4$D$_{1/2}$ &  -   &   6586884 &   6601400 &   6646599 & \\ 
   96 &      2s$^{2}$ 2p$^{5}$ 3s 3d &   o &   $^2$F$_{5/2}$ &   \sout{6556000}   &   6591507 &   6593543 &   6651977 & \\ 
   97 &      2s$^{2}$ 2p$^{5}$ 3p 3d &   e &   $^4$D$_{3/2}$ &  -   &   6595119 &   6608991 &   6654887 & \\ 
   98 &      2s$^{2}$ 2p$^{5}$ 3p 3d &   e &   $^4$D$_{5/2}$ &  -   &   6607993 &   6620899 &   6667902 & \\ 
   99 &      2s$^{2}$ 2p$^{5}$ 3s 3d &   o &   $^2$P$_{3/2}$ &   \sout{6595000} &   6617260 &   6616740 &   6686516 & \\
      &                             &     &                &   {\bf 6620000?} &        &        &      \\

  100 &      2s$^{2}$ 2p$^{5}$ 3p 3d &   e &   $^4$D$_{7/2}$ &  -   &   6623086 &   6634662 &   6682013 & \\ 
  101 &      2s$^{2}$ 2p$^{5}$ 3p 3d &   e &   $^4$G$_{7/2}$ &  -   &   6629973 &   6640658 &   6685459 & \\ 
  102 &      2s$^{2}$ 2p$^{5}$ 3p 3d &   e &   $^4$G$_{9/2}$ &  -   &   6631373 &   6641641 &   6685281 & \\ 
  103 &      2s$^{2}$ 2p$^{5}$ 3p 3d &   e &   $^2$D$_{5/2}$ &  -   &   6634688 &   6645828 &   6692155 & \\ 
  104 &      2s$^{2}$ 2p$^{5}$ 3p 3d &   e &   $^4$G$_{11/2}$ &  -   &   6636164 &   6646042 &   6690008 & \\ 
  105 &      2s$^{2}$ 2p$^{5}$ 3p 3d &   e &   $^2$D$_{3/2}$ &  -   &   6639303 &   6650162 &   6696538 & \\ 
  106 &      2s$^{2}$ 2p$^{5}$ 3p 3d &   e &   $^4$G$_{5/2}$ &  -   &   6649416 &   6659988 &   6704606 & \\ 
  107 &      2s$^{2}$ 2p$^{5}$ 3p 3d &   e &   $^2$F$_{7/2}$ &  -   &   6650899 &   6660419 &   6704344 & \\ 
  108 &      2s$^{2}$ 2p$^{5}$ 3p 3d &   e &   $^2$P$_{1/2}$ &  -   &   6652928 &   6664075 &   6708793 & \\ 
  109 &      2s$^{2}$ 2p$^{5}$ 3p 3d &   e &   $^4$P$_{1/2}$ &  -   &   6661688 &   6673499 &   6721372 & \\ 
  110 &      2s$^{2}$ 2p$^{5}$ 3p 3d &   e &   $^2$G$_{7/2}$ &  -   &   6663503 &   6673065 &   6718402 & \\ 
  111 &      2s$^{2}$ 2p$^{5}$ 3p 3d &   e &   $^4$P$_{3/2}$ &  -   &   6666392 &   6678388 &   6725771 & \\ 
  112 &      2s$^{2}$ 2p$^{5}$ 3p 3d &   e &   $^4$F$_{9/2}$ &  -   &   6675064 &   6684395 &   6729992 & \\ 
  113 &      2s$^{2}$ 2p$^{5}$ 3p 3d &   e &   $^4$P$_{5/2}$ &  -   &   6676103 &   6686826 &   6736070 & \\ 
  114 &      2s$^{2}$ 2p$^{5}$ 3p 3d &   e &   $^4$S$_{3/2}$ &  -   &   6676694 &   6686155 &   6731171 & \\ 
  115 &      2s$^{2}$ 2p$^{5}$ 3p 3d &   e &   $^4$D$_{7/2}$ &  -   &   6678774 &   6688328 &   6733893 & \\ 
  116 &      2s$^{2}$ 2p$^{5}$ 3p 3d &   e &   $^4$D$_{5/2}$ &  -   &   6679065 &   6689006 &   6735168 & \\ 
  117 &      2s$^{2}$ 2p$^{5}$ 3p 3d &   e &   $^4$P$_{3/2}$ &  -   &   6686282 &   6696202 &   6742881 & \\ 
  118 &      2s$^{2}$ 2p$^{5}$ 3p 3d &   e &   $^4$F$_{3/2}$ &  -   &   6692732 &   6701502 &   6747267 & \\ 
  119 &      2s$^{2}$ 2p$^{5}$ 3p 3d &   e &   $^4$F$_{9/2}$ &  -   &   6693560 &   6701731 &   6747879 & \\ 
  120 &      2s$^{2}$ 2p$^{5}$ 3p 3d &   e &   $^4$D$_{5/2}$ &  -   &   6701323 &   6708959 &   6755032 & \\ 
  121 &      2s$^{2}$ 2p$^{5}$ 3p 3d &   e &   $^4$F$_{7/2}$ &  -   &   6701728 &   6710310 &   6756669 & \\ 
  122 &      2s$^{2}$ 2p$^{5}$ 3p 3d &   e &   $^2$F$_{5/2}$ &  -   &   6703862 &   6712468 &   6759091 & \\ 
  123 &      2s$^{2}$ 2p$^{5}$ 3p 3d &   e &   $^2$P$_{1/2}$ &  -   &   6706716 &   6716454 &   6762890 & \\ 
  124 &      2s$^{2}$ 2p$^{5}$ 3p 3d &   e &   $^4$D$_{7/2}$ &  -   &   6708705 &   6716199 &   6762469 & \\ 
  125 &      2s$^{2}$ 2p$^{5}$ 3p 3d &   e &   $^2$D$_{3/2}$ &  -   &   6716985 &   6725873 &   6772014 & \\ 
  126 &      2s$^{2}$ 2p$^{5}$ 3p 3d &   e &   $^4$F$_{5/2}$ &  -   &   6719573 &   6727671 &   6774524 & \\ 
  127 &      2s$^{2}$ 2p$^{5}$ 3p 3d &   e &   $^2$D$_{3/2}$ &  -   &   6721823 &   6729273 &   6779084 & \\ 
  128 &      2s$^{2}$ 2p$^{5}$ 3p 3d &   e &   $^2$F$_{5/2}$ &  -   &   6727098 &   6735550 &   6776322 & \\ 
  129 &      2s$^{2}$ 2p$^{5}$ 3p 3d &   e &   $^4$D$_{1/2}$ &  -   &   6735806 &   6741753 &   6791670 & \\ 
  130 &      2s$^{2}$ 2p$^{5}$ 3p 3d &   e &   $^2$F$_{5/2}$ &  -   &   6736348 &   6743324 &   6789725 & \\ 
  131 &      2s$^{2}$ 2p$^{5}$ 3p 3d &   e &   $^2$S$_{1/2}$ &  -   &   6741107 &   6748420 &   6798866 & \\ 
  132 &      2s$^{2}$ 2p$^{5}$ 3p 3d &   e &   $^2$D$_{3/2}$ &  -   &   6741255 &   6749251 &   6794915 & \\ 
  133 &      2s$^{2}$ 2p$^{5}$ 3p 3d &   e &   $^4$F$_{7/2}$ &  -   &   6748758 &   6756215 &   6796906 & \\ 
  134 &      2s$^{2}$ 2p$^{5}$ 3p 3d &   e &   $^2$F$_{5/2}$ &  -   &   6753395 &   6760746 &   6802029 & \\ 
  135 &      2s$^{2}$ 2p$^{5}$ 3p 3d &   e &   $^2$F$_{7/2}$ &  -   &   6762854 &   6770538 &   6812665 & \\ 
  136 &      2s$^{2}$ 2p$^{5}$ 3p 3d &   e &   $^2$G$_{9/2}$ &  -   &   6765819 &   6771871 &   6818324 & \\ 
  137 &      2s$^{2}$ 2p$^{5}$ 3p 3d &   e &   $^4$D$_{3/2}$ &  -   &   6765828 &   6771678 &   6815497 & \\ 
  138 &      2s$^{2}$ 2p$^{5}$ 3p 3d &   e &   $^2$G$_{9/2}$ &  -   &   6768948 &   6772828 &   6826856 & \\ 
  139 &      2s$^{2}$ 2p$^{5}$ 3p 3d &   e &   $^4$F$_{3/2}$ &  -   &   6774024 &   6780566 &   6827080 & \\ 
  140 &      2s$^{2}$ 2p$^{5}$ 3p 3d &   e &   $^4$P$_{5/2}$ &  -   &   6780018 &   6784892 &   6837175 & \\ 
  141 &      2s$^{2}$ 2p$^{5}$ 3p 3d &   e &   $^4$P$_{1/2}$ &  -   &   6789682 &   6795392 &   6847309 & \\ 
  142 &      2s$^{2}$ 2p$^{5}$ 3p 3d &   e &   $^2$F$_{7/2}$ &  -   &   6790910 &   6794465 &   6849062 & \\ 
  143 &      2s$^{2}$ 2p$^{5}$ 3p 3d &   e &   $^2$D$_{5/2}$ &  -   &   6794015 &   6799158 &   6844589 & \\ 
  144 &      2s$^{2}$ 2p$^{5}$ 3p 3d &   e &   $^2$D$_{5/2}$ &  -   &   6799905 &   6804620 &   6854237 & \\ 
  145 &      2s$^{2}$ 2p$^{5}$ 3p 3d &   e &   $^2$P$_{3/2}$ &  -   &   6800872 &   6806274 &   6850264 & \\ 
  146 &      2s$^{2}$ 2p$^{5}$ 3p 3d &   e &   $^4$D$_{3/2}$ &  -   &   6805430 &   6812168 &   6855813 & \\ 
  147 &      2s$^{2}$ 2p$^{5}$ 3p 3d &   e &   $^4$D$_{1/2}$ &  -   &   6805751 &   6813392 &   6855558 & \\ 
  148 &      2s$^{2}$ 2p$^{5}$ 3p 3d &   e &   $^2$F$_{7/2}$ &  -   &   6809574 &   6812846 &   6865703 & \\ 
  149 &      2s$^{2}$ 2p$^{5}$ 3p 3d &   e &   $^2$P$_{3/2}$ &  -   &   6809716 &   6814230 &   6858541 & \\ 
  150 &      2s$^{2}$ 2p$^{5}$ 3p 3d &   e &   $^4$F$_{5/2}$ &  -   &   6811249 &   6815936 &   6861713 & \\

  151 &      2s$^{2}$ 2p$^{5}$ 3p 3d &   e &   $^2$D$_{3/2}$ &   {\bf 6831000?} &   6833056 &   6831282 &   6894915 & \\ 
  152 &      2s$^{2}$ 2p$^{5}$ 3p 3d &   e &   $^2$D$_{5/2}$ &   {\bf 6837100?} &   6837436 &   6838045 &   6893483 & \\

  153 &      2s$^{2}$ 2p$^{5}$ 3p 3d &   e &   $^2$P$_{1/2}$ &  -   &   6840857 &   6841603 &   6897847 & \\ 
  154 &      2s$^{2}$ 2p$^{5}$ 3p 3d &   e &   $^2$P$_{3/2}$ &  -   &   6859775 &   6859383 &   6918774 & \\ 
  155 &         2s 2p$^{6}$ 3s$^{2}$ &   e &   $^2$S$_{1/2}$ &  -   &   6869901 &   6861675 &  -   & \\ 
  156 &      2s$^{2}$ 2p$^{5}$ 3p 3d &   e &   $^2$P$_{1/2}$ &  -   &   6883420 &   6878831 &   6941928 & \\ 
  157 &      2s$^{2}$ 2p$^{5}$ 3p 3d &   e &   $^2$G$_{7/2}$ &  -   &   6883442 &   6884423 &   6936104 & \\ 
  158 &      2s$^{2}$ 2p$^{5}$ 3p 3d &   e &   $^2$D$_{5/2}$ &  -   &   6894168 &   6888882 &   6952863 & \\ 
  159 &      2s$^{2}$ 2p$^{5}$ 3p 3d &   e &   $^2$P$_{3/2}$ &  -   &   6897773 &   6894592 &   6954817 & \\ 
  160 &      2s$^{2}$ 2p$^{5}$ 3p 3d &   e &   $^2$S$_{1/2}$ &  -   &   6915372 &   6913088 &   6972752 & \\ 
  161 &      2s$^{2}$ 2p$^{5}$ 3p 3d &   e &   $^2$D$_{3/2}$ &  -   &   6928727 &   6920217 &   6995240 & \\ 
  162 &      2s$^{2}$ 2p$^{5}$ 3p 3d &   e &   $^2$D$_{5/2}$ &  -   &   6935337 &   6926090 &   7004293 & \\ 
  163 &   2s$^{2}$ 2p$^{5}$ 3d$^{2}$ &   o &   $^4$D$_{1/2}$ &  -   &   7025231 &   7037203 &  -   & \\ 
  164 &   2s$^{2}$ 2p$^{5}$ 3d$^{2}$ &   o &   $^4$D$_{3/2}$ &  -   &   7028016 &   7039419 &  -   & \\ 
  165 &   2s$^{2}$ 2p$^{5}$ 3d$^{2}$ &   o &   $^4$D$_{5/2}$ &  -   &   7032672 &   7043009 &  -   & \\ 
  166 &   2s$^{2}$ 2p$^{5}$ 3d$^{2}$ &   o &   $^4$D$_{7/2}$ &  -   &   7040652 &   7049523 &  -   & \\ 
  167 &   2s$^{2}$ 2p$^{5}$ 3d$^{2}$ &   o &   $^4$G$_{11/2}$ &  -   &   7048994 &   7055594 &  -   & \\ 
  168 &   2s$^{2}$ 2p$^{5}$ 3d$^{2}$ &   o &   $^4$G$_{9/2}$ &  -   &   7050157 &   7058175 &  -   & \\ 
  169 &   2s$^{2}$ 2p$^{5}$ 3d$^{2}$ &   o &   $^4$G$_{7/2}$ &  -   &   7055533 &   7064132 &  -   & \\ 
  170 &   2s$^{2}$ 2p$^{5}$ 3d$^{2}$ &   o &   $^2$F$_{5/2}$ &  -   &   7055676 &   7064629 &  -   & \\ 
  171 &            2s 2p$^{6}$ 3s 3p &   o &   $^4$P$_{1/2}$ &  -   &   7064005 &   7058900 &  -   & \\ 
  172 &   2s$^{2}$ 2p$^{5}$ 3d$^{2}$ &   o &   $^4$P$_{5/2}$ &  -   &   7066685 &   7074576 &  -   & \\ 
  173 &   2s$^{2}$ 2p$^{5}$ 3d$^{2}$ &   o &   $^4$F$_{9/2}$ &  -   &   7067562 &   7073220 &  -   & \\ 
  174 &   2s$^{2}$ 2p$^{5}$ 3d$^{2}$ &   o &   $^2$G$_{7/2}$ &  -   &   7068720 &   7075094 &  -   & \\ 
  175 &   2s$^{2}$ 2p$^{5}$ 3d$^{2}$ &   o &   $^2$D$_{3/2}$ &  -   &   7069932 &   7078209 &  -   & \\ 
  176 &            2s 2p$^{6}$ 3s 3p &   o &   $^4$P$_{3/2}$ &  -   &   7070546 &   7065086 &  -   & \\ 
  177 &   2s$^{2}$ 2p$^{5}$ 3d$^{2}$ &   o &   $^2$D$_{5/2}$ &  -   &   7076759 &   7081804 &  -   & \\ 
  178 &   2s$^{2}$ 2p$^{5}$ 3d$^{2}$ &   o &   $^4$P$_{3/2}$ &  -   &   7078608 &   7086971 &  -   & \\ 
  179 &            2s 2p$^{6}$ 3s 3p &   o &   $^4$P$_{5/2}$ &  -   &   7083684 &   7078148 &  -   & \\ 
  180 &   2s$^{2}$ 2p$^{5}$ 3d$^{2}$ &   o &   $^2$P$_{1/2}$ &  -   &   7086134 &   7092704 &  -   & \\ 
  181 &   2s$^{2}$ 2p$^{5}$ 3d$^{2}$ &   o &   $^2$H$_{11/2}$ &  -   &   7087677 &   7091469 &  -   & \\ 
  182 &   2s$^{2}$ 2p$^{5}$ 3d$^{2}$ &   o &   $^2$F$_{7/2}$ &  -   &   7094562 &   7099854 &  -   & \\ 
  183 &   2s$^{2}$ 2p$^{5}$ 3d$^{2}$ &   o &   $^4$D$_{5/2}$ &  -   &   7094609 &   7101485 &  -   & \\ 
  184 &   2s$^{2}$ 2p$^{5}$ 3d$^{2}$ &   o &   $^4$F$_{3/2}$ &  -   &   7097492 &   7103859 &  -   & \\ 
  185 &   2s$^{2}$ 2p$^{5}$ 3d$^{2}$ &   o &   $^4$D$_{7/2}$ &  -   &   7098186 &   7104169 &  -   & \\ 
  186 &   2s$^{2}$ 2p$^{5}$ 3d$^{2}$ &   o &   $^4$P$_{1/2}$ &  -   &   7099272 &   7104271 &  -   & \\ 
  187 &   2s$^{2}$ 2p$^{5}$ 3d$^{2}$ &   o &   $^2$G$_{9/2}$ &  -   &   7114444 &   7118468 &  -   & \\ 
  188 &   2s$^{2}$ 2p$^{5}$ 3d$^{2}$ &   o &   $^4$G$_{5/2}$ &  -   &   7117849 &   7124454 &  -   & \\ 
  189 &   2s$^{2}$ 2p$^{5}$ 3d$^{2}$ &   o &   $^2$D$_{3/2}$ &  -   &   7124454 &   7128734 &  -   & \\ 
  190 &            2s 2p$^{6}$ 3s 3p &   o &   $^2$P$_{1/2}$ &  -   &   7127309 &   7116914 &  -   & \\ 

 191 &   2s$^{2}$ 2p$^{5}$ 3d$^{2}$ &   o &   $^2$G$_{7/2}$ &   {\bf  7135000?}  &   7130233 &   7134361 &  -   & \\ 
  192 &   2s$^{2}$ 2p$^{5}$ 3d$^{2}$ &   o &   $^4$F$_{5/2}$ &  -   &   7136863 &   7142053 &  -   & \\ 
193 &            2s 2p$^{6}$ 3s 3p &   o &   $^2$P$_{3/2}$ &  -   &   7138849 &   7128949 &  -   & \\ 

  194 &   2s$^{2}$ 2p$^{5}$ 3d$^{2}$ &   o &   $^4$S$_{3/2}$ &  -   &   7145140 &   7148076 &  -   & \\ 
  195 &   2s$^{2}$ 2p$^{5}$ 3d$^{2}$ &   o &   $^4$F$_{7/2}$ &  -   &   7159683 &   7163088 &  -   & \\ 
  196 &   2s$^{2}$ 2p$^{5}$ 3d$^{2}$ &   o &   $^4$D$_{1/2}$ &  -   &   7160005 &   7165078 &  -   & \\ 
  197 &   2s$^{2}$ 2p$^{5}$ 3d$^{2}$ &   o &   $^2$G$_{9/2}$ &  -   &   7169545 &   7172806 &  -   & \\ 
  198 &   2s$^{2}$ 2p$^{5}$ 3d$^{2}$ &   o &   $^2$P$_{3/2}$ &  -   &   7181008 &   7183471 &  -   & \\ 
  199 &   2s$^{2}$ 2p$^{5}$ 3d$^{2}$ &   o &   $^2$F$_{7/2}$ &  -   &   7182632 &   7185214 &  -   & \\ 

  200 &   2s$^{2}$ 2p$^{5}$ 3d$^{2}$ &   o &   $^2$F$_{5/2}$ &  {\bf  7180800?} &   7183602 &   7186088 &  -   & \\ 
  201 &   2s$^{2}$ 2p$^{5}$ 3d$^{2}$ &   o &   $^2$F$_{5/2}$ &  {\bf  7191000?} &   7188617 &   7193832 &  -   & \\ 

 202 &   2s$^{2}$ 2p$^{5}$ 3d$^{2}$ &   o &   $^2$S$_{1/2}$ &  -   &   7200182 &   7199268 &  -   & \\ 
  203 &   2s$^{2}$ 2p$^{5}$ 3d$^{2}$ &   o &   $^4$D$_{3/2}$ &  -   &   7200476 &   7203616 &  -   & \\ 
  204 &   2s$^{2}$ 2p$^{5}$ 3d$^{2}$ &   o &   $^2$H$_{9/2}$ &  -   &   7207565 &   7209123 &  -   & \\ 
  205 &   2s$^{2}$ 2p$^{5}$ 3d$^{2}$ &   o &   $^2$D$_{5/2}$ &  -   &   7210871 &   7211689 &  -   & \\ 
  206 &            2s 2p$^{6}$ 3s 3p &   o &   $^2$P$_{1/2}$ &  -   &   7212169 &   7199268 &  -   & \\ 
  207 &   2s$^{2}$ 2p$^{5}$ 3d$^{2}$ &   o &   $^2$P$_{3/2}$ &  -   &   7214311 &   7211685 &  -   & \\ 
  208 &            2s 2p$^{6}$ 3s 3p &   o &   $^2$P$_{3/2}$ &  -   &   7216411 &   7203904 &  -   & \\ 
  209 &   2s$^{2}$ 2p$^{5}$ 3d$^{2}$ &   o &   $^2$P$_{1/2}$ &  -   &   7238076 &   7235908 &  -   & \\

  210 &   2s$^{2}$ 2p$^{5}$ 3d$^{2}$ &   o &   $^2$F$_{7/2}$ &  {\bf 7236000?} &   7241831 &   7242818 &  -   & \\ 
  211 &   2s$^{2}$ 2p$^{5}$ 3d$^{2}$ &   o &   $^2$D$_{5/2}$ &  {\bf 7240000?} &   7251131 &   7246119 &  -   & \\ 
  212 &   2s$^{2}$ 2p$^{5}$ 3d$^{2}$ &   o &   $^2$D$_{3/2}$ &  -   &   7257952 &   7253388 &  -   & \\ 
  213 &   2s$^{2}$ 2p$^{5}$ 3d$^{2}$ &   o &   $^2$P$_{3/2}$ &  {\bf 7266000?} &   7269242 &   7263947 &  -   & \\ 

  214 &   2s$^{2}$ 2p$^{5}$ 3d$^{2}$ &   o &   $^2$P$_{1/2}$ &  -   &   7309222 &   7303627 &  -   & \\ 
  215 &         2s 2p$^{6}$ 3p$^{2}$ &   e &   $^4$P$_{1/2}$ &  -   &   7384058 &   7374540 &  -   & \\ 
  216 &         2s 2p$^{6}$ 3p$^{2}$ &   e &   $^2$D$_{5/2}$ &  -   &   7389574 &   7381183 &  -   & \\ 
  217 &         2s 2p$^{6}$ 3p$^{2}$ &   e &   $^2$D$_{3/2}$ &  -   &   7390566 &   7382311 &  -   & \\ 
  218 &         2s 2p$^{6}$ 3p$^{2}$ &   e &   $^4$P$_{3/2}$ &  -   &   7395633 &   7386567 &  -   & \\ 
  219 &         2s 2p$^{6}$ 3p$^{2}$ &   e &   $^4$P$_{5/2}$ &  -   &   7411114 &   7401807 &  -   & \\ 
  220 &         2s 2p$^{6}$ 3p$^{2}$ &   e &   $^2$P$_{1/2}$ &  -   &   7418366 &   7408049 &  -   & \\ 
  221 &         2s 2p$^{6}$ 3p$^{2}$ &   e &   $^2$P$_{3/2}$ &  -   &   7435885 &   7425663 &  -   & \\ 
  222 &            2s 2p$^{6}$ 3s 3d &   e &   $^4$D$_{1/2}$ &  -   &   7480669 &   7473130 &  -   & \\ 
  223 &            2s 2p$^{6}$ 3s 3d &   e &   $^4$D$_{3/2}$ &  -   &   7481725 &   7473674 &  -   & \\ 
  224 &            2s 2p$^{6}$ 3s 3d &   e &   $^4$D$_{5/2}$ &  -   &   7483510 &   7474659 &  -   & \\ 
  225 &            2s 2p$^{6}$ 3s 3d &   e &   $^4$D$_{7/2}$ &  -   &   7486075 &   7476203 &  -   & \\ 
  226 &         2s 2p$^{6}$ 3p$^{2}$ &   e &   $^2$S$_{1/2}$ &  -   &   7506071 &   7491289 &  -   & \\ 
  227 &            2s 2p$^{6}$ 3s 3d &   e &   $^2$D$_{3/2}$ &  -   &   7552884 &   7537978 &  -   & \\ 
  228 &            2s 2p$^{6}$ 3s 3d &   e &   $^2$D$_{5/2}$ &  -   &   7556269 &   7540221 &  -   & \\ 
  229 &            2s 2p$^{6}$ 3s 3d &   e &   $^2$D$_{3/2}$ &  -   &   7600455 &   7583866 &  -   & \\ 
  230 &            2s 2p$^{6}$ 3s 3d &   e &   $^2$D$_{5/2}$ &  -   &   7600984 &   7583934 &  -   & \\ 
  231 &      2s$^{2}$ 2p$^{5}$ 3s 4s &   o &   $^4$P$_{5/2}$ &  -   &   7655895 &   7668872 &  -   & \\ 
  232 &      2s$^{2}$ 2p$^{5}$ 3s 4s &   o &   $^4$P$_{3/2}$ &  -   &   7667243 &   7679557 &  -   & \\ 
  233 &      2s$^{2}$ 2p$^{5}$ 3s 4s &   o &   $^2$P$_{1/2}$ &  -   &   7677217 &   7689062 &  -   & \\ 
  234 &      2s$^{2}$ 2p$^{5}$ 3s 4s &   o &   $^2$P$_{3/2}$ &  -   &   7691383 &   7701732 &  -   & \\ 
  235 &            2s 2p$^{6}$ 3p 3d &   o &   $^4$F$_{3/2}$ &  -   &   7716155 &   7709523 &  -   & \\ 
  236 &            2s 2p$^{6}$ 3p 3d &   o &   $^4$F$_{5/2}$ &  -   &   7721250 &   7713928 &  -   & \\ 
  237 &            2s 2p$^{6}$ 3p 3d &   o &   $^4$F$_{7/2}$ &  -   &   7729551 &   7721465 &  -   & \\ 
  238 &            2s 2p$^{6}$ 3p 3d &   o &   $^4$F$_{9/2}$ &  -   &   7740363 &   7731661 &  -   & \\ 
  239 &            2s 2p$^{6}$ 3p 3d &   o &   $^2$D$_{5/2}$ &  -   &   7749958 &   7740486 &  -   & \\ 
  240 &            2s 2p$^{6}$ 3p 3d &   o &   $^2$D$_{3/2}$ &  -   &   7753184 &   7743954 &  -   & \\ 

  \hline
\end{longtable}